\documentclass[useAMS,usenatbib,usegraphics]{mn2e}
\usepackage{ifpdf} 
\pdfoutput=1
\usepackage{graphicx}
\usepackage{latexsym,amssymb}
\usepackage{amsmath}
\usepackage{gensymb}
\usepackage{capt-of}
\usepackage{textpos}
\usepackage{url}
\usepackage{multirow}
\usepackage{color}
\usepackage[T1]{fontenc}
\usepackage{longtable}

\usepackage{hyperref}
    \hypersetup{
        bookmarks=true,         % show bookmarks bar?
        unicode=false,          % non-Latin characters in Acrobat�s bookmarks
        pdftoolbar=true,        % show Acrobat�s toolbar?
        pdfmenubar=true,        % show Acrobat�s menu?
        pdffitwindow=false,     % window fit to page when opened
        pdfstartview={FitH},    % fits the width of the page to the window
        pdftitle={My title},    % title
        pdfauthor={Author},     % author
        pdfsubject={Subject},   % subject of the document
        pdfcreator={Creator},   % creator of the document
        pdfproducer={Producer}, % producer of the document
        pdfkeywords={keyword1} {key2} {key3}, % list of keywords
        pdfnewwindow=true,      % links in new window
        colorlinks=false,       % false: boxed links; true: colored links
        linkcolor=black,          % color of internal links (change box color with linkbordercolor)
        citecolor=blue,        % color of links to bibliography
        filecolor=black,      % color of file links
        urlcolor=black,           % color of external links
        linkbordercolor={1 0 0},
        citebordercolor={0 1 0}
    }

\makeatletter
\Hy@AtBeginDocument{%
  \def\@pdfborder{0 0 1}% Overrides border definition set with colorlinks=true
  \def\@pdfborderstyle{/S/U/W 0}% Overrides border style set with colorlinks=true
}
\makeatother

\newcommand{\kms}{km\,s$^{-1}$}

\newcommand{\Msun}{M$_\odot$}

\newcommand{\MLsun}{$\Upsilon_\odot$}

\newcommand{\Reff}{\ensuremath{R_e}}

\newcommand{\MLstar}{\ensuremath{\Upsilon_\star}}

\newcommand{\Mbh}{\ensuremath{M_\mathrm{\bullet}}}
\newcommand{\Mstar}{\ensuremath{M_{\star}}}
\newcommand{\hst}{\textit{HST}}                     % HST
\newcommand{\ppak}{\textit{PPAK}}                     % PPAK
\newcommand{\sersic}{S\'{e}rsic}		%S�rsic

\title[The massive dark halo of NGC\,1281]{The massive dark halo of the compact, early-type galaxy NGC\,1281}
\author[A. Y{\i}ld{\i}r{\i}m et al.]
{\parbox{\textwidth}{Ak{\i}n Y{\i}ld{\i}r{\i}m$^{1}$\thanks{E-mail: yildirim@mpia.de},
Remco C. E. van den Bosch$^{1}$,
Glenn van de Ven$^{1}$,
Aaron Dutton$^{1}$,
Ronald L\"{a}sker$^{1}$,
Bernd Husemann$^{2}$,
Jonelle L. Walsh$^{3}$,
Karl Gebhardt$^{4}$,
Kayhan G\"ultekin$^{5}$
and Ignacio Mart\'{i}n-Navarro$^{6,7}$}\vspace{0.4cm}\\
\parbox{\textwidth}{
$^{1}$Max Planck Institute for Astronomy, K\"onigstuhl 17, 69117 Heidelberg, Germany\\
$^{2}$European Southern Observatory, Karl-Schwarzschild-Str. 2, 85748 Garching, Germany\\
$^{3}$George P. and Cynthia Woods Mitchell Institute for Fundamental Physics and Astronomy, Department of Physics and Astronomy, Texas A\&M University, College Station, TX 77843, USA\\
$^{4}$Department of Astronomy, The University of Texas at Austin, Austin, TX 78712, USA\\
$^{5}$Department of Astronomy, University of Michigan, Ann Arbor, MI 48109, USA\\
$^{6}$Instituto de Astrof\'{i}sica de Canaris, E-38200 La Laguna, Tenerife, Spain\\
$^{7}$Departamento de Astrof\'{i}sica, Universidad de La Laguna, E-38205 La Laguna, Tenerife, Spain}}

\begin{document}

%\date{In prep. for MNRAS}
\date{Accepted 2015 November 10. Received November 9; in original form 2015 August 31}

%\pagerange{\pageref{firstpage}--\pageref{lastpage}} \pubyear{2002}

\def\LaTeX{L\kern-.36em\raise.3ex\hbox{a}\kern-.15em
    T\kern-.1667em\lower.7ex\hbox{E}\kern-.125emX}

\maketitle

\label{firstpage}

\begin{abstract}
We investigate the compact, early-type galaxy NGC\,1281 with integral field unit observations to map the stellar line-of-sight velocity distribution (LOSVD) out to 5 effective radii and construct orbit-based dynamical models to constrain its dark and luminous matter content. Under the assumption of mass-follows-light, the \textit{H}-band stellar mass-to-light ratio (M/L) is \MLstar\ = 2.7$\pm$0.1 \MLsun, and higher than expected from our stellar population synthesis fits with either a canonical Kroupa (\MLstar\ = 1.3 \MLsun) or Salpeter (\MLstar\ = 1.7 \MLsun) stellar initial mass function. Such models also cannot reproduce the details of the LOSVD. Models with a dark halo recover the kinematics well and indicate that NGC\,1281 is dark matter dominated, making up $\sim$ 90 per cent of the total enclosed mass within the kinematic bounds. Parameterised as a spherical NFW profile, the dark halo mass is 11.5 $\le$ log($M_{DM}$/$M_{\scriptstyle \odot}$) $\le$ 11.8 and the stellar M/L is 0.6 $\le$ \MLstar/\MLsun\ $\le$ 1.1. However, this M/L is lower than predicted by its old stellar population. Moreover, the halo mass within the kinematic extent is ten times larger than expected based on $\Lambda$CDM predictions, and an extrapolation yields cluster sized dark halo masses. Adopting \MLstar\ = 1.7 \MLsun\ yields more moderate dark halo virial masses, but these models fit the kinematics worse. A non-NFW model might solve the discrepancy between the unphysical consequences of the best-fitting dynamical models and models based on more reasonable assumptions for the dark halo and stellar mass-to-light ratio, which are disfavoured according to our parameter estimation.

\end{abstract}


%============================= section 1 =============================

\section{Introduction}
\label{sec:introduction}

In the cold dark matter paradigm ($\Lambda$CDM), galaxies form via cooling of gas in the potential wells of extended, virialised dark matter halos. While this picture has been very successful in describing the local, large scale structures of the universe \citep{2007ApJS..170..377S}, conflicts arise between the predictions and observations on galaxy scales as the resolution of cosmological $N$-body simulations steadily increases. The poster child of these lingering discrepancies is the "core-cusp" problem where the central dark matter density distribution of dwarf- and low surface brightness galaxies indicate a cored-like behaviour \citep{1994Natur.370..629M,1998ApJ...499...41M}, in contrast to simulations which show a clear cusp \citep{1996ApJ...462..563N}. But, the Galaxy, too, stubbornly refuses to provide evidence of a significant amount of massive dark subhalos, commonly known as the "Missing Satellites Problem" \citep{1993MNRAS.264..201K,1999ApJ...522...82K,1999ApJ...524L..19M,2011MNRAS.415L..40B}.\\

Efforts to reconcile both models and observations have focused on complex baryonic feedback processes on the theoretical side. Issues on the observational side, however, are manifold. Although proof for the presence of dark matter in galaxies dates back to the 80s, based on extended neutral gas discs that map the dark halo at many effective radii (\Reff) in late-type galaxies \citep{1980ApJ...238..471R,1985ApJ...295..305V}, the detection of halos in early type galaxies (ETGs) is not as straightforward due to the lack of similar features. Different tracers have therefore been employed - e.g. planetary nebulae \citep{2007ApJ...664..257D,2009MNRAS.394.1249C}, globular clusters \citep{2003ApJ...591..850C} and hot X-ray gas \citep{2006ApJ...646..899H} - but the dynamical modelling of each of these brings along its own set of assumptions and leads to partially conflicting answers \citep[e.g.][]{2003Sci...301.1696R,2006MNRAS.366.1253P}. The bulk of work has therefore focused on the analysis of stellar kinematics with two-integral axisymmetric Jeans models or state-of-the-art orbit-based dynamical models which are more conclusive and indicate that dark matter is a key ingredient \citep{1997ApJ...488..702R,2001AJ....121.1936G,2006MNRAS.366.1126C,2007MNRAS.382..657T,2009MNRAS.396.1132T,2009MNRAS.398..561W,2013MNRAS.432.1709C}. The spatial coverage of the stellar kinematics, however, is usually limited to regions where the stellar mass is assumed to dominate the total mass budget, and the attempt to break the degeneracy between the stellar and dark mass in these regions is complicated. As a consequence, the halo parameterisation remains a matter of debate and the amount of dark matter strongly varies between no dark matter in some individual cases and a more typical dark matter contribution of up to $\sim$ 50 per cent to the total enclosed mass within one effective radius \citep[e.g.][]{2007ApJ...667..176G,2008ApJ...684..248B}.\\

In this work, we present orbit-based dynamical models of the compact ETG NGC\,1281. The data acquisition and analysis of NGC\,1281 has been carried out as part of a larger program that aims at the detailed study of compact, high stellar velocity dispersion galaxies \citep{2012Natur.491..729V,2013MNRAS.434L..31L,2015ApJS..218...10V,2015ApJ...808..183W,2015MNRAS.452.1792Y,2015ApJ...808...79F,WBGY}. The program's target sample was a result of the Hobby-Eberly Massive Galaxy Survey (HETMGS) which, based on the sphere of influence argument, looked for high velocity dispersion galaxies that are candidates for hosting (over-)massive supermassive black holes (SMBHs).

Interestingly, these compact objects also have striking similarities to the massive and quiescent galaxy population at higher redshifts \citep{2003ApJ...587L..79F,2003ApJ...587L..83V} - which exhibit high velocity dispersions \citep[$\sigma \ge$ 300 \kms;][]{2009Natur.460..717V,2011ApJ...736L...9V,2012ApJ...754....3T,2013ApJ...764L...8B,2013ApJ...771...85V}, have small effective radii \citep[\Reff\ $\le$ 2 kpc;][]{2005ApJ...626..680D,2006ApJ...650...18T,2007ApJ...656...66Z,2008ApJ...688...48V,2008ApJ...677L...5V,2012ApJ...749..121S,2014ApJ...788...28V} and disky surface brightness profiles \citep{2005ApJ...624L...9T,2011ApJ...730...38V,2013ApJ...762...83C} - and which are believed to constitute the cores of today's most massive ellipticals \citep{2010ApJ...709.1018V,2014ApJ...791...45V,2015arXiv150702291W}.

We yet lack high-spatial resolution spectroscopic observations of NGC\,1281 which will enable us to resolve the dynamics close to the SMBH, but complementary data sets of high-spatial resolution imaging and integral-field unit (IFU) observations are now available to study its internal structure and large scale mass distribution. In particular the wide-field IFU data presented here extends to a scale of $\sim$ 5 effective radii ($\simeq$ 7\,kpc) which should be sufficient for a dynamical constraint on the dark matter halo.\\

This paper is organised as follows: Section \hyperref[sec:data]{\ref{sec:data}} presents the data, consisting of deep \hst\ imaging and wide-field \ppak\ IFU stellar kinematics. In Section \hyperref[sec:dynamical_analysis]{\ref{sec:dynamical_analysis}}, we carry out orbit-based dynamical models of NGC\,1281 to constrain its stellar and dark content. We discuss the reliability and significance of our findings in Section \hyperref[sec:discussion]{\ref{sec:discussion}} and summarise our results in Section \hyperref[sec:summary]{\ref{sec:summary}}. Throughout this paper we adopt 5th year results of the Wilkinson Microwave Anisotropy Probe \citep[WMAP;][]{2009ApJS..180..225H}, with a Hubble constant of $H_0 = 70.5$ \kms\ Mpc$^{-1}$, a matter density of $\Omega_{M} = 0.27$ and a cosmological constant of $\Omega_{\lambda} = 0.73$.

%============================= section 2 =============================
\section{Data}
\label{sec:data}
%=====================================================================

%---------------------------------------------------------------------
\subsection{HST IMAGING}
\label{sec:imaging}
%---------------------------------------------------------------------

For brevity, we refer to \cite{2015MNRAS.452.1792Y} and the references therein, which covers in detail the data acquisition and reduction pipelines. Here, we confine ourselves to only a brief description of the main steps.\\

Photometric data of NGC\,1281 have been obtained by the \textit{HST} WFC3 in \textit{H}-band (F160W), as part of program GO: 13050 (PI: van den Bosch), resulting in a total of three dithered full- and four subarray exposures with a total integration time of $\sim$ 1400s. The purpose of the 1.7s short subarray exposures is to mitigate the effect of saturation in the nucleus of the long 450s fullarrays.
The seven flat-field calibrated exposures are reduced and combined via \textsc{Astrodrizzle} \citep{2012drzp.book.....G} to obtain a super-sampled image with a resolution of 0.06\arcsec/pixel, while correcting for geometric and photometric distortions.

For the photometric analysis, we make use of the point-spread function (PSF) of the Cosmic Assembly Near-Infrared Deep Extragalactic Survey \citep[CANDELS;][]{2012ApJS..203...24V}. The PSF is created with \textsc{TinyTim} \citep{1995ASPC...77..349K} in the F160W filter and 'drizzled' to replicate the final PSF at the desired plate scale of 0.06\arcsec.\\

\begin{figure*}
		\begin{center}
		\includegraphics[width=.49\textwidth]{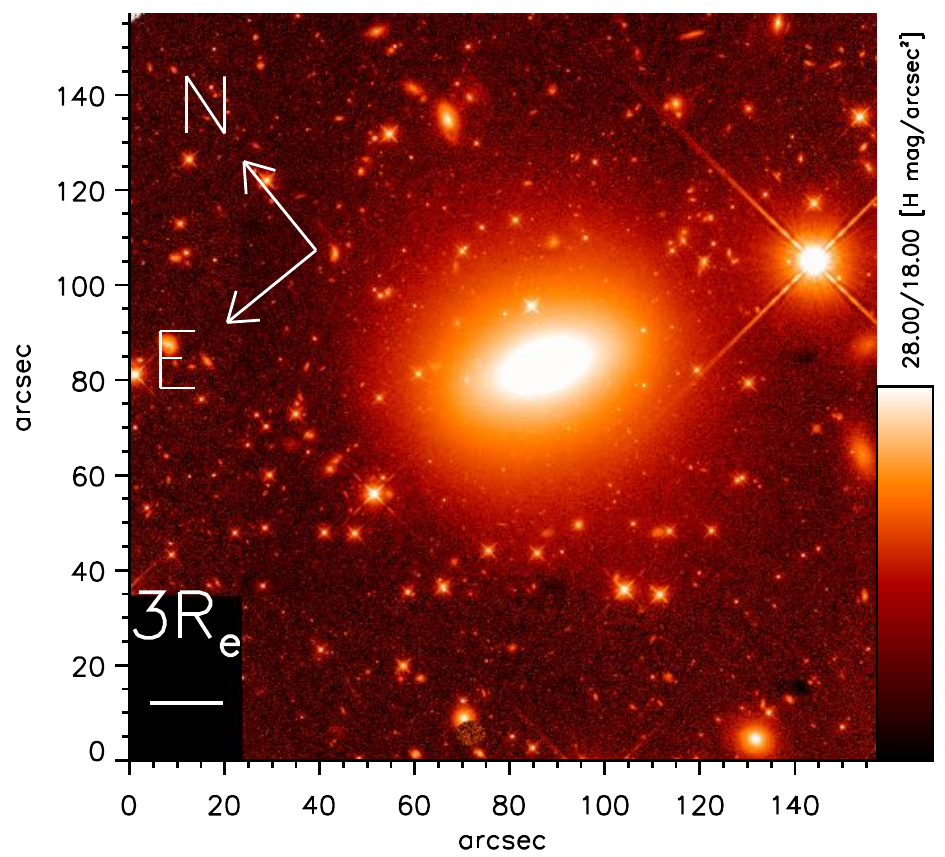}
		\includegraphics[width=.49\textwidth]{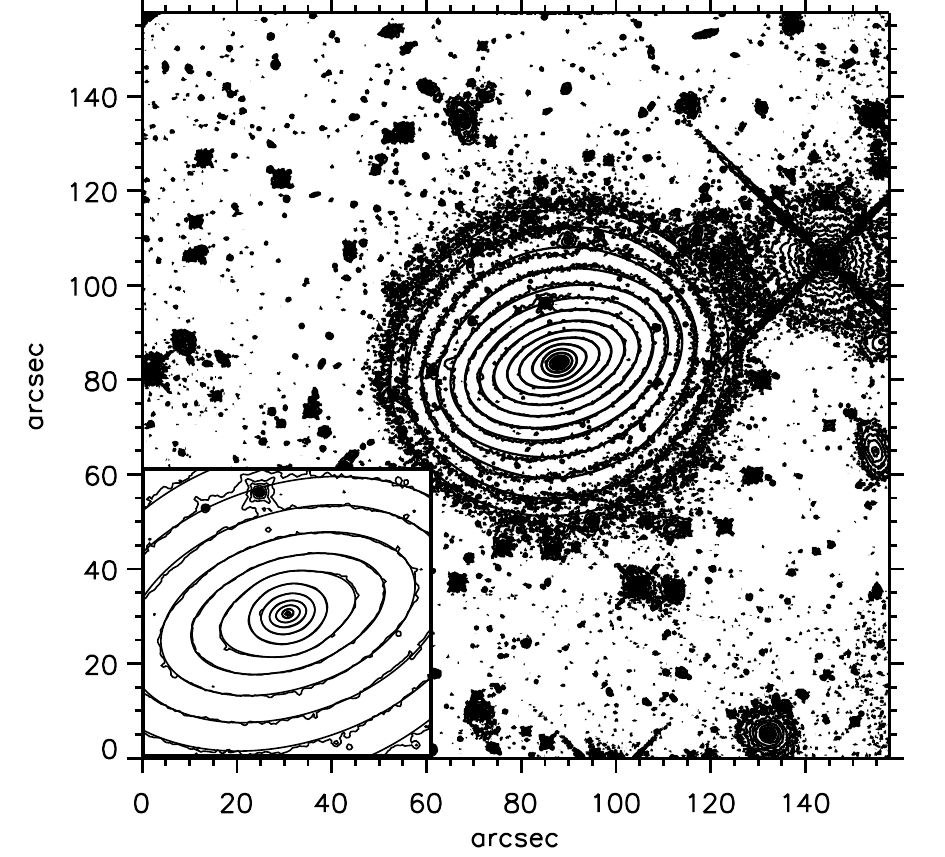}
		\end{center}
	\caption{\textit{Left:} \textit{HST} \textit{H}-band image of NGC\,1281, which covers a field of 150 $\square$\arcsec, with a final scale of 0.06\arcsec/pixel. \textit{Right:} Contour map of the same image. The MGE contours are over-plotted in black. The bottom left plot shows the accurate reproduction of the surface brightness profile within the central 30 arcsec$^{2}$.}
	\label{fig:ngc1281_phot}
\end{figure*}

NGC\,1281 is a compact ETG in the Perseus cluster with an E5 morphological classification according to the NASA/IPAC Extragalactic Database (NED). Redshift measurements of this object place it at a Hubble flow distance of $60\pm1$\,Mpc, at which 1\arcsec\ corresponds to an absolute scale of 0.29\,kpc. A circular aperture that contains half of the light has a size of $R_{1/2}$ = 4.5\arcsec\ (i.e. 1.3\,kpc). The galaxy has a total apparent magnitude of $m_{H,Vega}$ = 10.50 \footnote{All magnitudes presented in this paper are corrected for galactic extinction (0.085 mag) based on SDSS re-calibrated values \citep{2011ApJ...737..103S} of infrared dust maps \citep{1998ApJ...500..525S}.}. Assuming a stellar M/L of 1.7 \MLsun\ in \textit{H}-band, as indicated by its uniformly old stellar population (see Sec. \hyperref[sec:ngc1281_imf]{\ref{sec:ngc1281_imf}}) and adopting an absolute magnitude for the Sun of 3.32, this implies a total stellar mass of log(\Mstar) = 10.92.

We further investigate its photometric properties by carrying out a single and two \sersic\ decomposition of the high-resolution \textit{H}-band imaging while generously masking any contamination by fore- and background objects. A single \sersic\ fit has an apparent magnitude of $m_{H,Vega}$ = 10.36 with an effective (semi major axis) radius of 6.8\arcsec\ (i.e. 2\,kpc), a projected axis ratio ($q=b/a$) of 0.63 and a single \sersic\ index of $n$ = 3.9. Residuals of this fit are strong and show an overestimation of the outer and inner surface brightness (SB) profile. The residuals also reveal the presence of a central dust disk with an apparent minor to major axis ratio of 0.35 and a semi-major axis length of 0.3\,kpc. If this intrinsically flat disk traces the position angle (PA) of its host, we can infer the inclination of this galaxy to be 70\degree. A fit with two \sersic\ components improves the fit considerably. Here, a rather flat ($q = 0.7$), inner (\Reff = 1.7\arcsec), 'bulge'-like component ($n = 2.2$) with a magnitude of $m_{H,Vega}$ = 11.61 is embedded in a flat ($q = 0.6$), outer (\Reff = 10\arcsec), disk-like ($n=1.3$) component with a magnitude of $m_{H,Vega}$ = 10.97. The fit has a bulge-to-total ratio ($B/T$) of 0.55.\\

\begin{table}
	\caption{Multi-Gaussian-Expansion of NGC\,1281's \textit{HST} (F160W) \textit{H}-band image. The columns display the number of each Gaussian, beginning with the innermost one (i), their total \textit{H}-band magnitude, corrected for galactic extinction (0.085 mag) (ii), their effective radius (iii) and their corresponding \sersic\ index (iv) as well as their apparent flattening (v). The MGE has been obtained by fixing the PA of all Gaussians to 69\degree, with the image aligned N.-E. An absolute magnitude for the Sun of 3.32 in \textit{H}-band is employed to convert the values into stellar surface densities.}
	\begin{center}
	\begin{tabular}{ c  c  c  c  c }
		\hline
		\# of components & mag [H, Vega] & \Reff\ [arcsec] & $n$ & $q$ \\
		(i) & (ii) & (iii) & (iv) & (v) \\
		\hline
		1 & 14.59  &    0.15    & 0.5 &  0.58 \\
		2 & 15.57  &    0.40    & 0.5 &  0.99 \\
		3 & 13.20  &    0.78   &  0.5 &  0.60 \\
		4 & 12.93  &    1.49   &  0.5 &  0.65 \\
		5 & 13.31  &    1.83   &  0.5 &  0.99 \\
		6 & 12.15  &    5.04  &  0.5  & 0.52 \\
		7 & 11.89  &    9.64  &  0.5 & 0.59 \\
		8 & 12.51  &    15.71  &  0.5  & 0.90 \\
		\hline
	\end{tabular}
	\vspace{2ex}
	\label{tab:ngc1281_mge}
	\end{center}
\end{table}

To obtain a stellar mass model, we parameterise it's surface brightness profile by a set of multiple Gaussians \citep[MGE;][]{1992A&A...253..366M,1994A&A...285..723E}. In our case, the MGE consists of 8 Gaussians with a common centre and a fixed position angle (PA) of 69\degree, measured counter-clockwise from the y-axis to the galaxy major axis, with the image aligned N.-E. (i.e. north is up and east is left).

\begin{table*}
	\caption{Bi-symmetrised \ppak\ stellar kinematics of NGC\,1281. The columns indicate the bin number (i), the location of the bin centroids in x (ii) and y (iii) direction, the LOSVD consisting of the mean line-of-sight velocity $v$ (iv), the  velocity dispersion $\sigma$ (vi), the higher order Gauss-Hermite moments $h_3$ (viii) and $h_4$ (x) and the corresponding 1$\sigma$ errors. This table will be published in its entirety in the electronic edition of the journal. A portion is shown here for guidance.}
	\begin{center}
	\begin{tabular}{ c  c  c  c  c  c  c  c  c  c  c}
		\hline
		 bin \# & x [arcsec] & y [arcsec] & $v$ [\kms] & $\Delta v$ [\kms] & $\sigma$ [\kms] & $\Delta\sigma$ [\kms] & $h_3$ & $\Delta h_3$ & $h_4$ & $\Delta h_4$ \\
		 (i) & (ii) & (iii) & (iv) & (v) & (vi) & (vii) & (viii) & (ix) & (x) & (xi) \\
		\hline
		1 & 0.532 & 0.213 & -44.386 &  4.833 & 266.280 & 6.346 & 0.022   & 0.015 & 0.035 & 0.019 \\
		2 & 0.532 & 1.213 & -17.088 &  5.198 & 260.249 & 6.933 & 0.001 & 0.017 & 0.021 & 0.023 \\
		3 & -0.468 & 0.213 &  54.905  & 4.789 & 266.233 & 6.228 &-0.029  & 0.015  & 0.036 & 0.019 \\
		.. & ..  &    ..   &  .. &  ..  & .. & .. & .. & .. & & \\
		167 & 17.229 & 2.834 & -123.053 & 21.557 & 169.807 & 28.480 & 0.036 & 0.121 & 0.040 & 0.129 \\
		\hline
	\end{tabular}
	\vspace{2ex}
	\label{tab:ngc1281_ppak}
	\end{center}
\end{table*}

In Fig. \hyperref[fig:ngc1281_phot]{\ref{fig:ngc1281_phot}}, we illustrate the deep \hst\ \textit{H}-band image (left) and the MGE parameterisation (right). The MGE accurately reproduces the SB profile of this object throughout the whole radial extent. There is no position angle twist in the photometric profile and the PA varies by less than 3\degree\ between the inner- and outermost isophotes.\\

%---------------------------------------------------------------------
\subsection{PPAK KINEMATICS}
\label{sec:kinematics}
%---------------------------------------------------------------------

\begin{figure}
		\begin{center}
		\includegraphics[width=.47\textwidth]{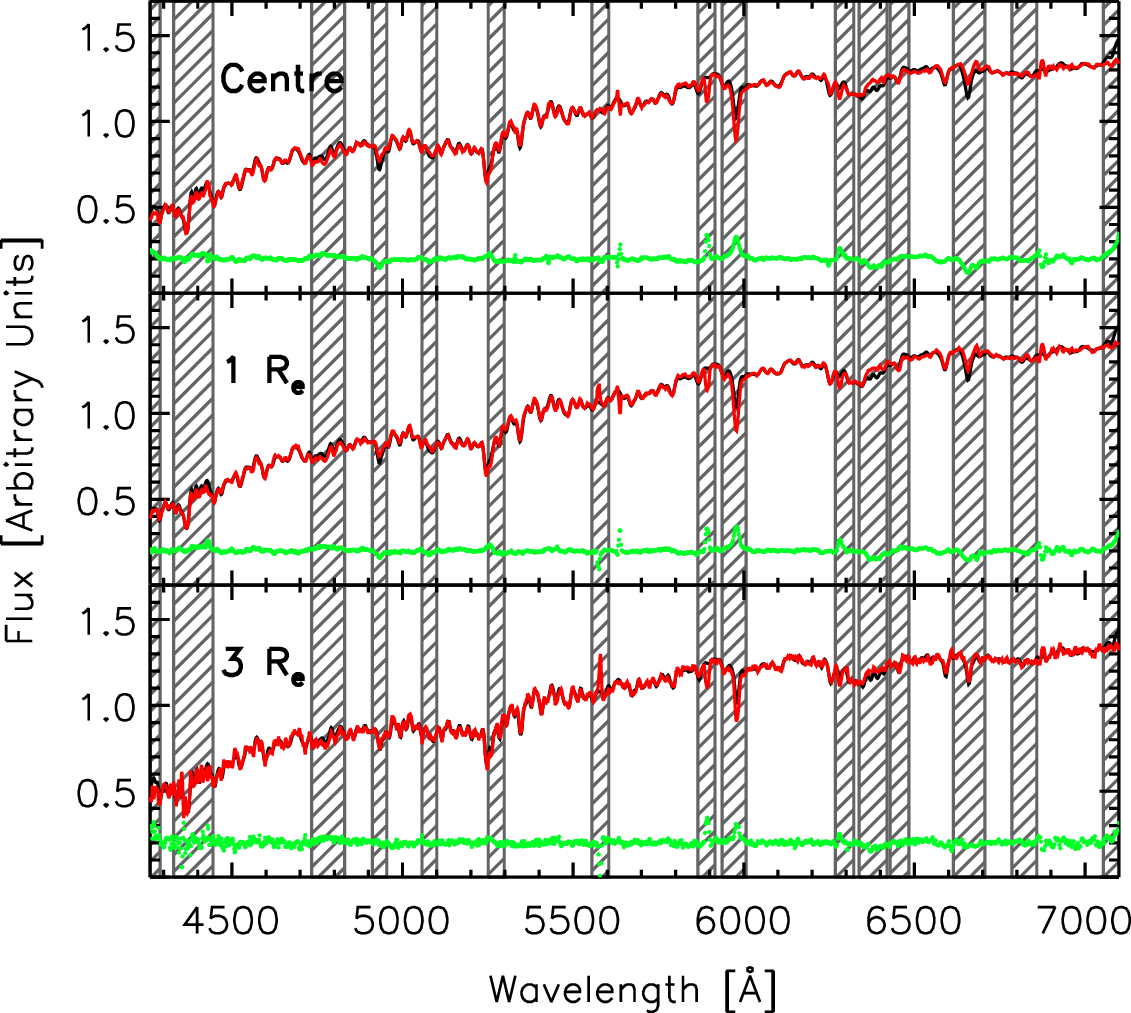}
		\end{center}
	\caption{\textit: The \ppak\ IFU spectra (black), the stellar templates convolved with the best-fitting LOSVD (red) and the model residuals (green) of NGC\,1281, for the bin in the centre (top) and for bins at 1 (middle) and 3 \Reff (bottom). The flux and residuals have been shifted by an arbitrary amount. The grey shaded boxes depict the spectral masks.}
	\label{fig:ngc1281_ppxf}
\end{figure}

\begin{figure*}
		\begin{center}
		\includegraphics[width=.99\textwidth]{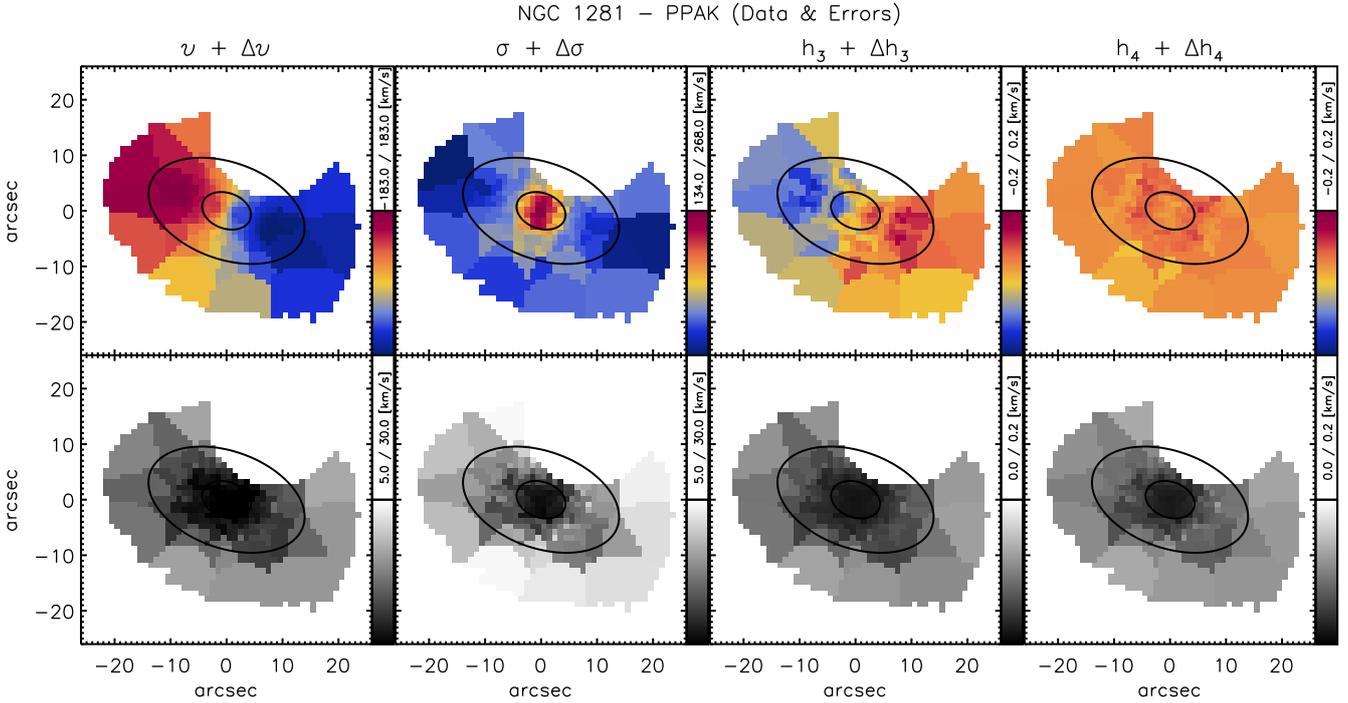}
		\end{center}
	\caption{\textit{Top}: PPAK IFU stellar kinematic maps of NGC\,1281, showing the mean line-of-sight velocity $v$, velocity dispersion $\sigma$, and Gauss-Hermite moments $h_3$ and $h_4$. The maps show a fast rotation around the short axis of 183 \kms\ and a central velocity dispersion of 268 \kms. Overplotted are contours of constant surface brightness at 1 and 3 \Reff, as measured from the deep \hst\ photometry. The empty region above the centre is generously masked due to the presence of a bright foreground star. \textit{Bottom}: Corresponding uncertainty maps. Maps are oriented N.-E., i.e. north is up and east is left.}
	\label{fig:ngc1281_kin}
\end{figure*}

Kinematic data of NGC\,1281 have been obtained at the 3.5m telescope at Calar Alto, with the \textit{Potsdam Multi Aperture Spectrograph} (Roth et al. 2005) in the \ppak\ mode \citep{2004AN....325..151V,2006PASP..118..129K}, making use of the low-resolution V500 grating. This setup has a resolving power of \textit{R} = 850, at 5400 \AA\ and a spectral resolution of 6.3 \AA\ across the nominal 3745 - 7500 \AA\ spectral range, corresponding to a velocity resolution of $\sigma$ =150 \kms. The \ppak\ IFU consists of 382 2.7\arcsec diameter fibres, which are bundled to a hexagonal shape and cover a field of view (FOV) of $1.3\arcmin \times 1.3\arcmin$. 331 of the total 382 fibres are science fibres that provide a 100 per cent covering factor with a 3 dither-pointing strategy while 36 fibres, located 72\arcsec away from the centre, are used to sample the sky.

The observing strategy was defined by the overall goal to acquire wide-field stellar kinematics out to several effective radii. This was accomplished on January 5 and 6, 2013 with a total exposure time of 6 hours on source, achieved by 3 complete runs with 3 dither pointings of 2400s each, divided into 2 frames \`{a} 1200s.

The data reduction follows the dedicated reduction pipeline of the Calar Alto Legacy Integral Field Spectroscopic Area (\textsc{CALIFA}) Survey \citep{2012A&A...538A...8S,2013A&A...549A..87H} and includes bias and straylight subtraction, cosmic ray rejection \citep{2012A&A...545A.137H}, optimal fibre extraction, fibre flat-fielding, flexure correction and wavelength and flux calibration using spectrophotometric standard stars. The sky spectrum is the mean of only the 30 faintest sky fibres - to exclude any contamination by stars or low SB brightness objects - and subtracted from its associated science exposure. The 3 dither-pointings are then resampled to the final data cube with a 1\arcsec\ sampling using a distance-weighted interpolation algorithm \citep{2012A&A...538A...8S}.\\

We extract the line-of-sight velocity distribution (LOSVD) by fitting the spatially binned spectra \citep{2003MNRAS.342..345C} with a non-negative, linear combination of stellar templates \citep{2004PASP..116..138C} from the Indo-U.S. stellar library \citep{2004ApJS..152..251V} in the vignetting and sensitivity limited useful spectral range of 4200-7000 \AA. The spectra were binned to reach a minimum S/N of 40 in each Voronoi zone, resulting in a total of 167 bins that cover the stellar kinematics of NGC\,1281 out to $\sim$ 24\arcsec\ (i.e. 5 effective radii or roughly 7\,kpc). The stellar templates have not been fixed during the fitting process, but are allowed to vary between spatial bins. The best-fitting stellar templates, though, are commonly dominated by cool (T$_{\rm eff} \le 5400$ K) K2IV, K1V, G0, M1III, M4III, K5 stars, with solar to sub-solar metallicities (see also Sec. \hyperref[sec:ngc1281_imf]{\ref{sec:ngc1281_imf}}, for a more thorough analysis of the stellar populations characteristics). Sky and emission line features have been generously masked beforehand, while ensuring that important stellar absorption features, such as H$\beta$, Mg $Ib$ and Fe 5015 are not affected by the spectral masks (Fig. \hyperref[fig:ngc1281_ppxf]{\ref{fig:ngc1281_ppxf}}). Moreover, 15th degree additive Legendre polynomials have been employed to correct the template continuum shapes during the fit. Finally, a series of tests have been carried out to asses the robustness of the extracted stellar kinematics, which show that variations in the stellar templates, spectral masks and degrees of polynomials yield consistent data sets at the 2$\sigma$ level and thus consistent modelling results (Sec. \hyperref[sec:caveats]{\ref{sec:caveats}}).

Fig. \hyperref[fig:ngc1281_kin]{\ref{fig:ngc1281_kin}} displays the bi-symmetrised, two-dimensional mean line-of-sight velocity $v$, the velocity dispersion $\sigma$ and the higher order Gauss-Hermite moments $h_3$ and $h_4$ per bin on the plane of the sky. The kinematic maps reveal an extended, fast and regular rotation around the short axis, with a maximum velocity of $\sim$ 183 \kms. The peak in velocity dispersion is $\sim$ 268 \kms\ and drops to $\sim$ 134 \kms\ for the outermost bins. When measuring the stellar velocity dispersion from the stacked spectra within a 3\arcsec\ aperture, we obtain a central velocity dispersion of 261$\pm$6 \kms, which is well in line with the SDSS literature value of 258$\pm$2 \kms. Furthermore, we observe a strong anti-correlation between $v$ and $h_3$, as is expected for axisymmetric systems \citep{2005ApJ...626..159B,2004AJ....127.3192C}, and throughout positive $h_4$ values within the whole radial extent. Errors on the kinematic moments have been determined via 100 Monte Carlo simulations, with the penalisation term set to zero, where random Gaussian noise was added to the spectrum. The errors are small, with mean values of 8 \kms\ and 14 \kms\ for $v$ and $\sigma$ and 0.05 and 0.06 for $h_3$ and $h_4$ respectively. The radial increase in the measurement errors is due to the rapid decline of NGC\,1281's surface brightness. The S/N of the individual spaxels drops rapidly and the outermost bins cannot accumulate enough spaxels to reach the target S/N.

To replicate the PSF of the \ppak\ observations, we fit a PSF convolved, reconstructed image to the MGE of the \hst\ high-resolution data. The PSF is expanded by multiple, round Gaussians and in our case the image is well reproduced by two components with a dispersion of 1.53\arcsec\ and 5.39\arcsec\ and corresponding weights of 84\% and 16\% respectively.

%============================= section 3 =============================
\section{Dynamical Analysis}
\label{sec:dynamical_analysis}
%=====================================================================

%---------------------------------------------------------------------
\subsection{Schwarzschild Models}
\label{sec:schwarzschild}
%---------------------------------------------------------------------

The dynamical analysis uses the triaxial implementation of Schwarzschild's orbit superposition method \citep{2008MNRAS.385..647V}. In contrast to Jeans models which suffer from the mass-anisotropy degeneracy \citep{1982MNRAS.200..361B,1987ApJ...313..121M}, Schwarzschild's numerical approach bypasses our ignorance of the orbital distribution and hence the orbital anisotropy by means of the information encoded in the higher order velocity moments \citep{1987MNRAS.224...13D,1993ApJ...407..525V,1993MNRAS.265..213G}. The implementation starts with a trial potential, including the contribution of luminous and dark matter, which determines the initial conditions for all supported orbit families. In our case, this representative orbit library, which thoroughly samples all integrals of motion, consists of 7533 orbits sampled along 31 logarithmically spaced equipotential shells out to 150\arcsec. Every shell is used as a starting point for 9 orbits in each radial and angular direction. The orbits are then integrated numerically and, during orbit integration, their intrinsic quantities are stored and projected onto the space of observables for comparison with the data. After PSF convolution, the models can match the binned LOSVD in a least-square sense by linearly adding up the contributions of the orbital building blocks. The linear superposition of orbits in a gravitational potential has to be non-negative, since the weights assigned to each orbit represent its mass. With the orbital weights being also a numerical representation of the distribution function \citep{1984ApJ...287..475V}, this ensures the distribution function to be positive, i.e. physically meaningful, everywhere. Self-consistency of the models is taken care of by demanding that the orbital weights are also capable of reproducing the intrinsic and projected stellar masses within an accuracy of 2 per cent, which reflects the uncertainties when parameterising the SB of galaxies with multiple Gaussians. Finally, the parameters defining the gravitational potential of the host are varied and the best-fitting models and parameter constraints are determined via a $\chi^2$ analysis, which allows us to infer the contributions of the underlying gravitational constituents 	to the total mass budget. The reliability of this approach has been tested extensively and proven to be a powerful method to infer the internal mass distribution and dynamical structure of galaxies \citep{2008MNRAS.385..614V,2009MNRAS.398.1117V,2010MNRAS.401.1770V}.\\

For our mass model, we consider three gravitational sources; the stellar mass \Mstar, which is the deprojected intrinsic luminosity density times a constant stellar-mass-to light ratio \MLstar, the black hole mass \Mbh\ of a central SMBH and the dark halo, parameterised by a spherically symmetric NFW profile \citep{1996ApJ...462..563N} with concentration $c_{DM}$ and total viral mass $M_{DM} = M_{200}$. The models cover a large range in parameters space of \MLstar/\MLsun\ $\in$ [0.4, 3.0], log(\Mbh/$M_{\scriptstyle \odot}$) $\in$ [8.0, 10.4], log($c_{DM}$) $\in$ [0.8, 1.2] and log($M_{DM}$/\Mstar) $\in$ [-2, 6], which is motivated by theoretical and observational constraints on the stellar mass-to-light ratio for stellar population synthesis (SPS) models \citep{1996ApJS..106..307V} with a Kroupa or Salpeter stellar initial mass function (IMF), the black hole mass from predictions of the black hole scaling relations \citep{2009ApJ...698..198G} and the dark halo parameters from numerical $N$-body simulations within the $\Lambda$CDM framework \citep{2001MNRAS.321..559B,2008MNRAS.391.1940M,2010ApJ...710..903M,2010MNRAS.407....2D}.\\

In this work we restrict ourselves to an oblate axisymmetric stellar system. The fast and regular rotation around its apparent short-axis, the negligible variation in its photometric PA and the results of shape inversions of a large sample of fast-rotating ETGs \citep{2014MNRAS.444.3340W}, show that axial symmetry is a well justified assumption of the intrinsic shape of NGC\,1281. Hence, the inclination is the only free viewing parameter that is needed to deproject the surface brightness and to infer the stellar mass density and hence the stellar gravitational potential of the galaxy.

The flattest Gaussian in our fit has an axis ratio of $q = 0.52$ (Table \hyperref[tab:ngc1281_mge]{\ref{tab:ngc1281_mge}}), which implies possible inclinations of 60\degree\ - 90\degree\ (with 90\degree\ being edge-on). In our case, though, the inclination is reasonably well constrained by the central dust disc to be 70\degree. Nevertheless, to assess the reliability of our results with respect to changes in the inclination, we have also probed a more edge-on view of 80\degree\ and found no significant deviations (< 5 per cent) for the parameter constraints that are reported below.

%---------------------------------------------------------------------
\subsection{Modelling Results}
\label{sec:ngc1281_schwarzschild}
%---------------------------------------------------------------------

\begin{figure}
		\begin{center}
		\includegraphics[width=.47\textwidth]{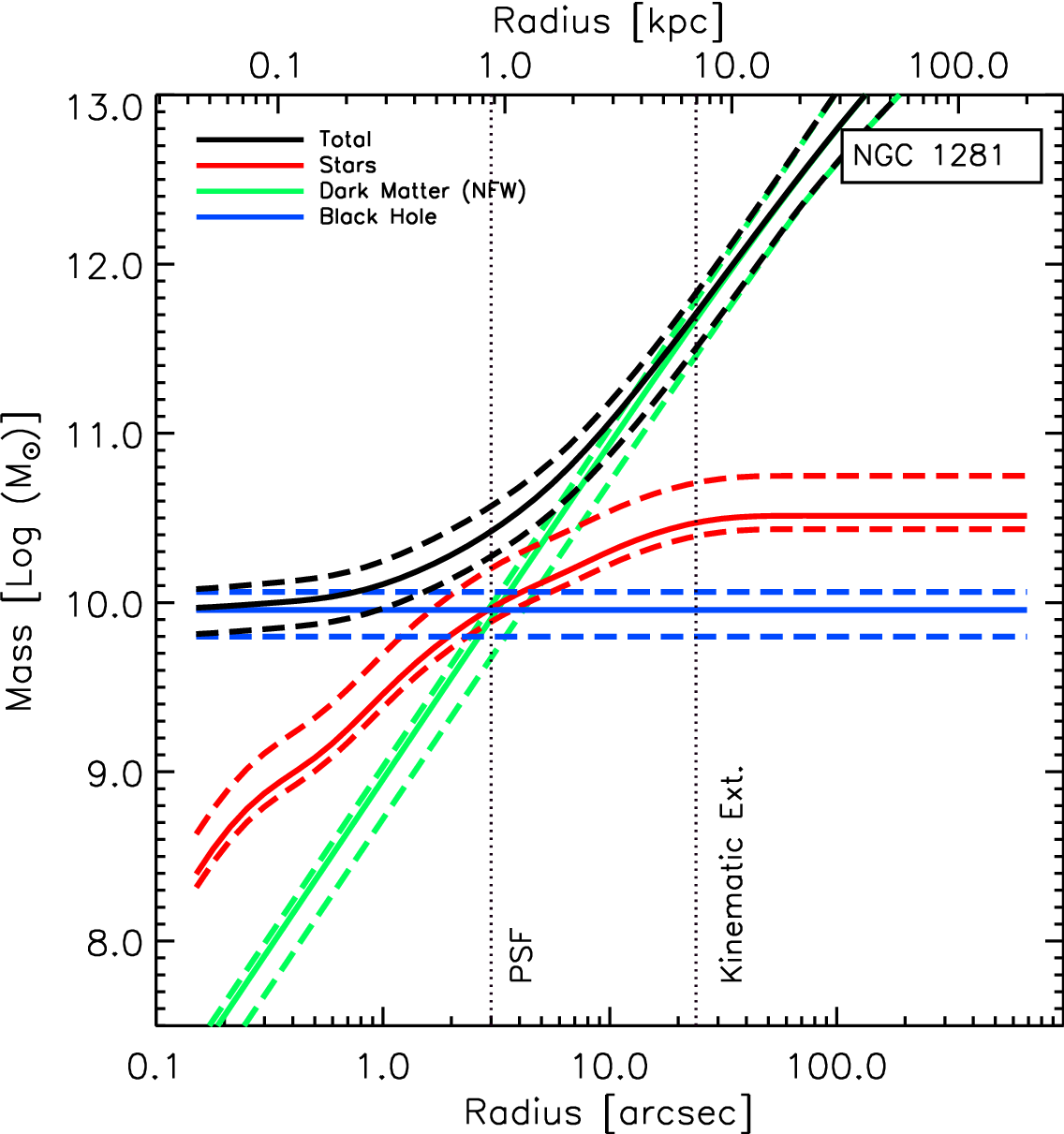}
		\end{center}
	\caption{Intrinsic, enclosed mass distribution of NGC\,1281 as a function of radius, obtained from our orbit-based dynamical models with a spherically symmetric NFW halo. Solid lines represent best-fitting values and dashed lines 3$\sigma$ confidence intervals. The dotted vertical lines indicate the resolution limit and the extent of the kinematic observations.}
	\label{fig:ngc1281_model_mass}
\end{figure}

\begin{figure*}
		\begin{center}
		\includegraphics[width=.975\textwidth]{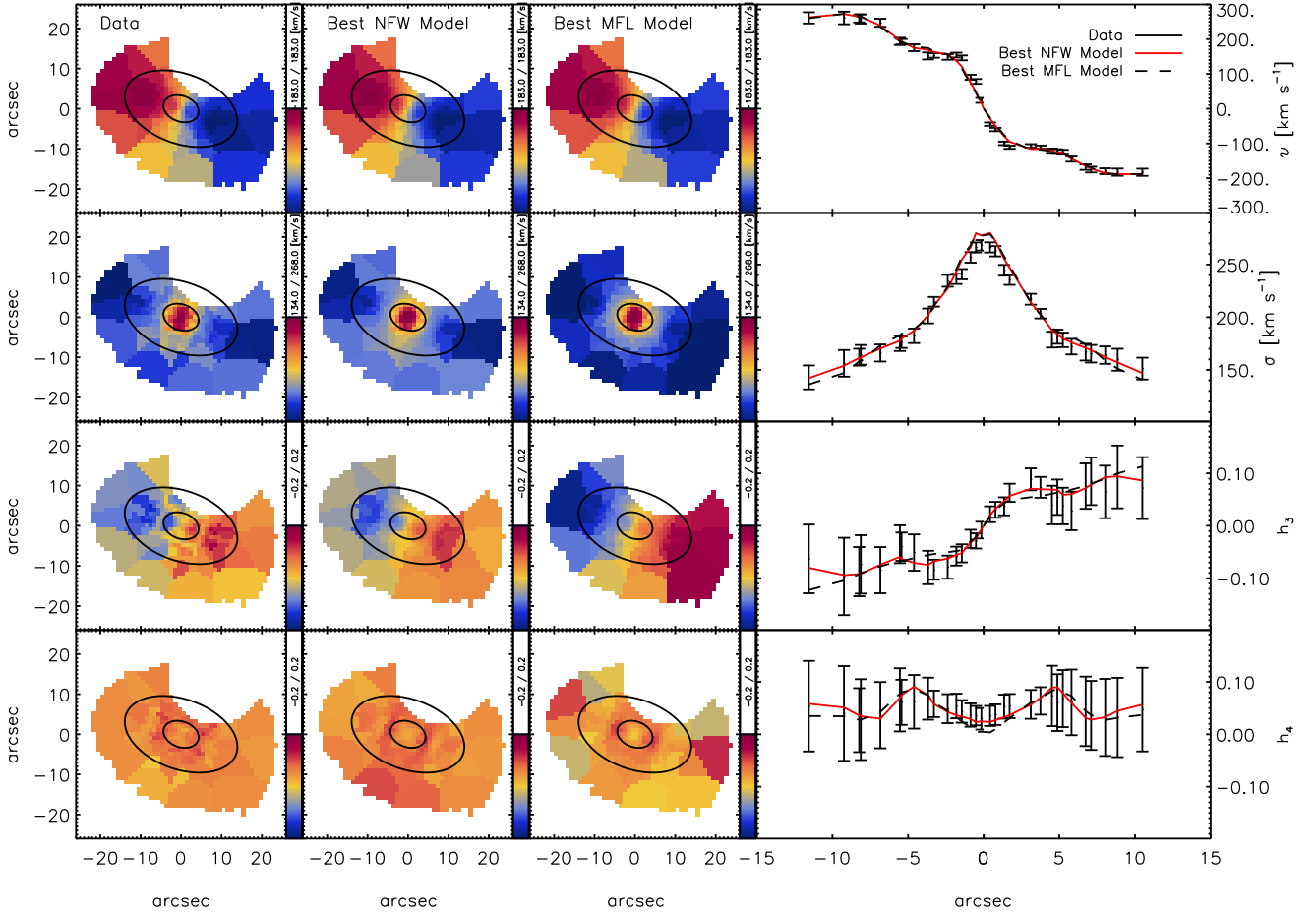}
		\end{center}
	\caption{\textit{First column}: \ppak\ IFU data consisting of the mean line-of-sight velocity $v$, velocity dispersion $\sigma$, $h_3$ and $h_4$. All maps are oriented N.-E., i.e. north is up and east is left. \textit{Second column}: Overall best-fitting Schwarzschild model predictions with a NFW halo and a reduced $\chi^2$ of 0.32. \textit{Third column}: Best-fitting model without a halo, i.e. mass-follows-light (MFL). Models without a dark component cannot recover the details of the LOSVD and deviate from the best-fitting dynamical models (which include a massive dark halo) by $\Delta\chi^2 \sim$ 114. \textit{Fourth column}: Data, best-fitting NFW model and best-fitting MFL model for the measurements where the Voronoi bin centroids fall within a 1\arcsec\ wide strip along the major axis. We point out that the last column needs to be interpreted carefully. The apparently small differences between both models can be misleading, as the most prominent deviations can be found in bins which do not lie along the major axis.}

	\label{fig:ngc1281_model_kin}
\end{figure*}

The results of our orbit-based dynamical models are presented in Fig. \hyperref[fig:ngc1281_model_mass]{\ref{fig:ngc1281_model_mass}}. The figure shows the total enclosed mass (black), the stellar mass (red), the dark matter content (green) and the black hole mass (blue) as a function of radius. The solid lines denote the values for the best-fitting model, with the dashed lines indicating the statistical 3$\sigma$ uncertainties for one degree of freedom. We obtain a stellar mass of log(\Mstar/$M_{\scriptstyle \odot}$) = 10.5$^{+0.2}_{-0.1}$, a dark halo to stellar mass ratio of log($M_{DM}$/\Mstar) = 3.6$^{+1.5}_{-0.5}$, a black hole mass of log(\Mbh/$M_{\scriptstyle \odot}$) = 10.0$^{+0.1}_{-0.2}$ and an \textit{H}-band stellar M/L of \MLstar/\MLsun\ = 0.7$^{+0.4}_{-0.1}$. At small radii ($\le$ 3\arcsec) the black hole is the dominant contributor to the total mass budget but quickly released from its role by the dark halo, which already takes over at $\sim$ 4.5\arcsec\ (i.e. 1 \Reff\ or roughly 1.3\,kpc). The total enclosed mass within the kinematic extent is log($M_{TOT}$/$M_{\scriptstyle \odot}$) = 11.7$^{+0.1}_{-0.2}$.

Clearly, models without a dark halo are excluded. We illustrate this in Fig. \hyperref[fig:ngc1281_model_kin]{\ref{fig:ngc1281_model_kin}}, where we present (from left to right) the data, the overall best-fitting model (log($c_{DM}$) = 1.2, log($M_{DM}/$\Mstar) = 3.6, log(\Mbh/$M_{\scriptstyle \odot}$) = 10.0 and \MLstar/\MLsun = 0.7) and the best-fitting model without a halo (\MLstar/\MLsun = 2.7). The difference between the overall best-fitting model with a halo and the best-fitting model without a halo is $\Delta\chi^2 \sim 114 $. Thus, a model without a halo is excluded at a significance of more than 5$\sigma$, with most of the $\chi^2$ difference attributable to the fit to $v, \sigma$ and $h_3$. \\

\begin{figure*}
		\centering
		\includegraphics[width=.24\linewidth]{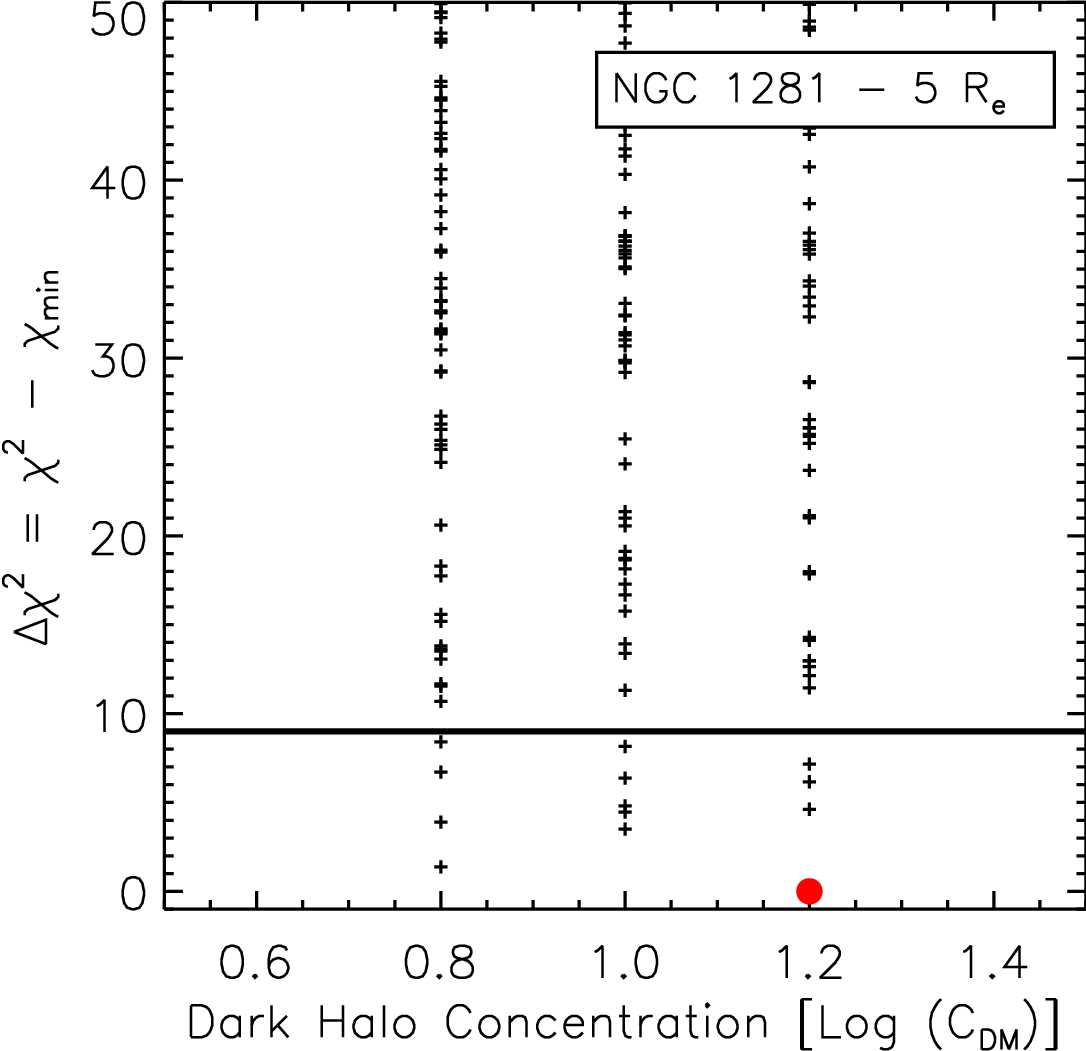}
		\vspace{4ex}
		\includegraphics[width=.24\linewidth]{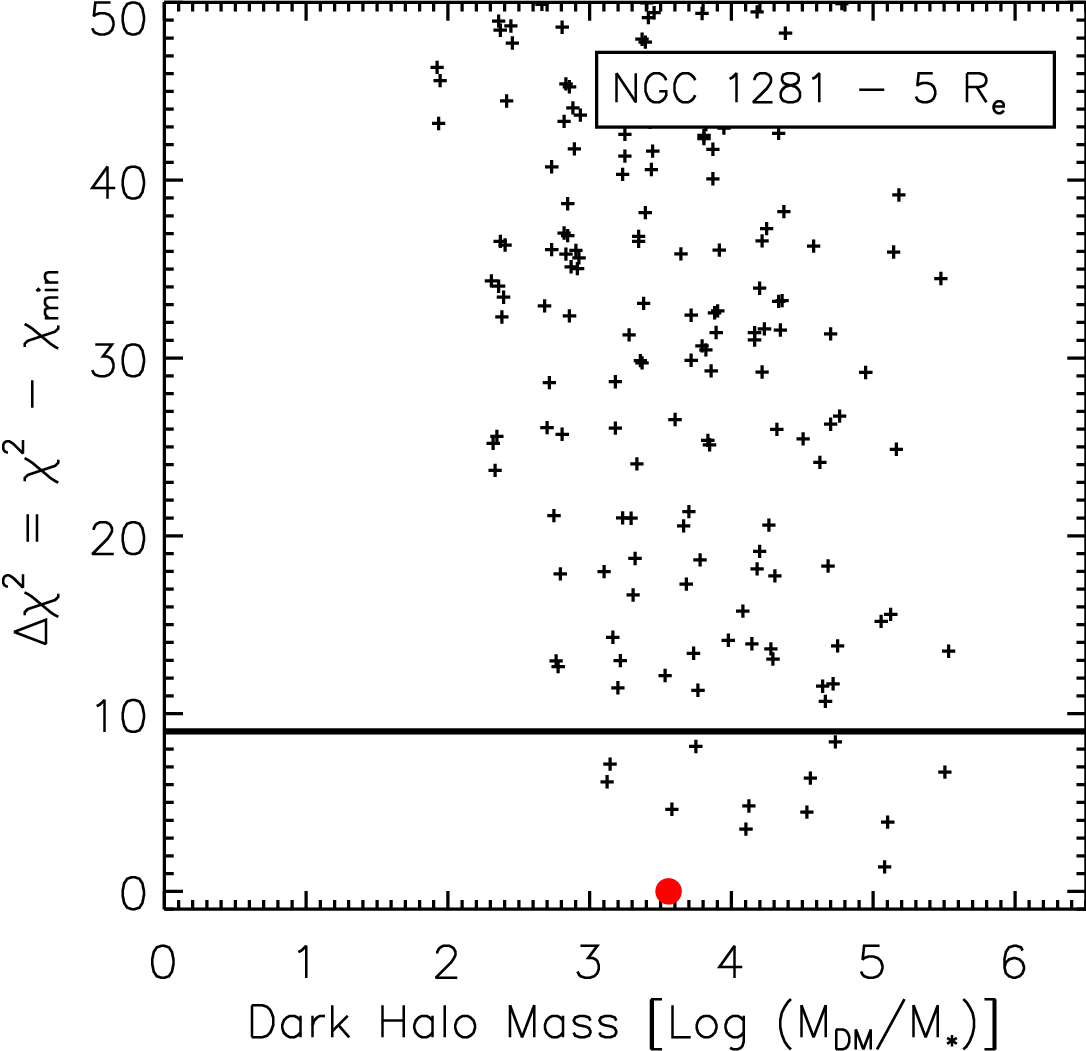}
		\includegraphics[width=.24\linewidth]{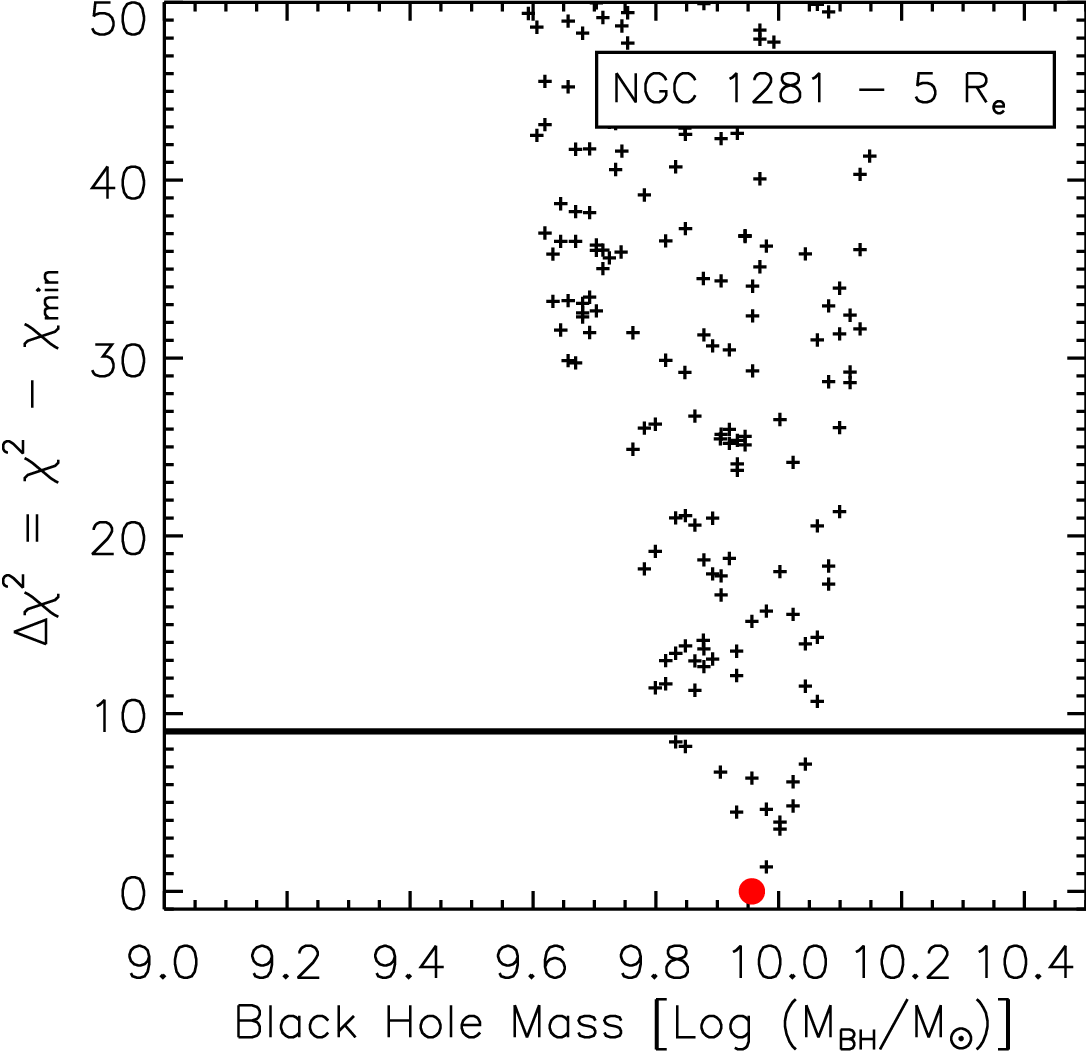}
		\includegraphics[width=.24\linewidth]{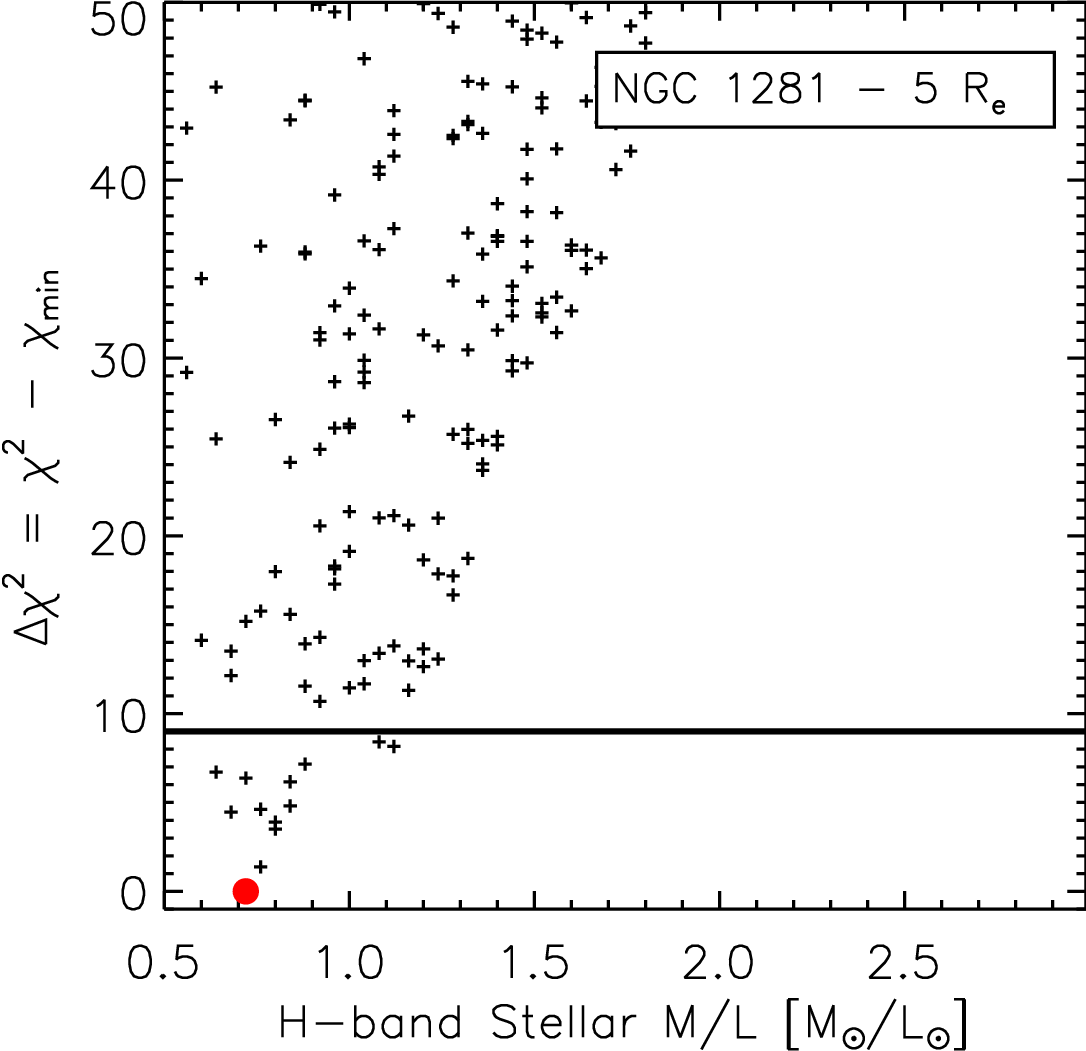}
		\includegraphics[width=.24\linewidth]{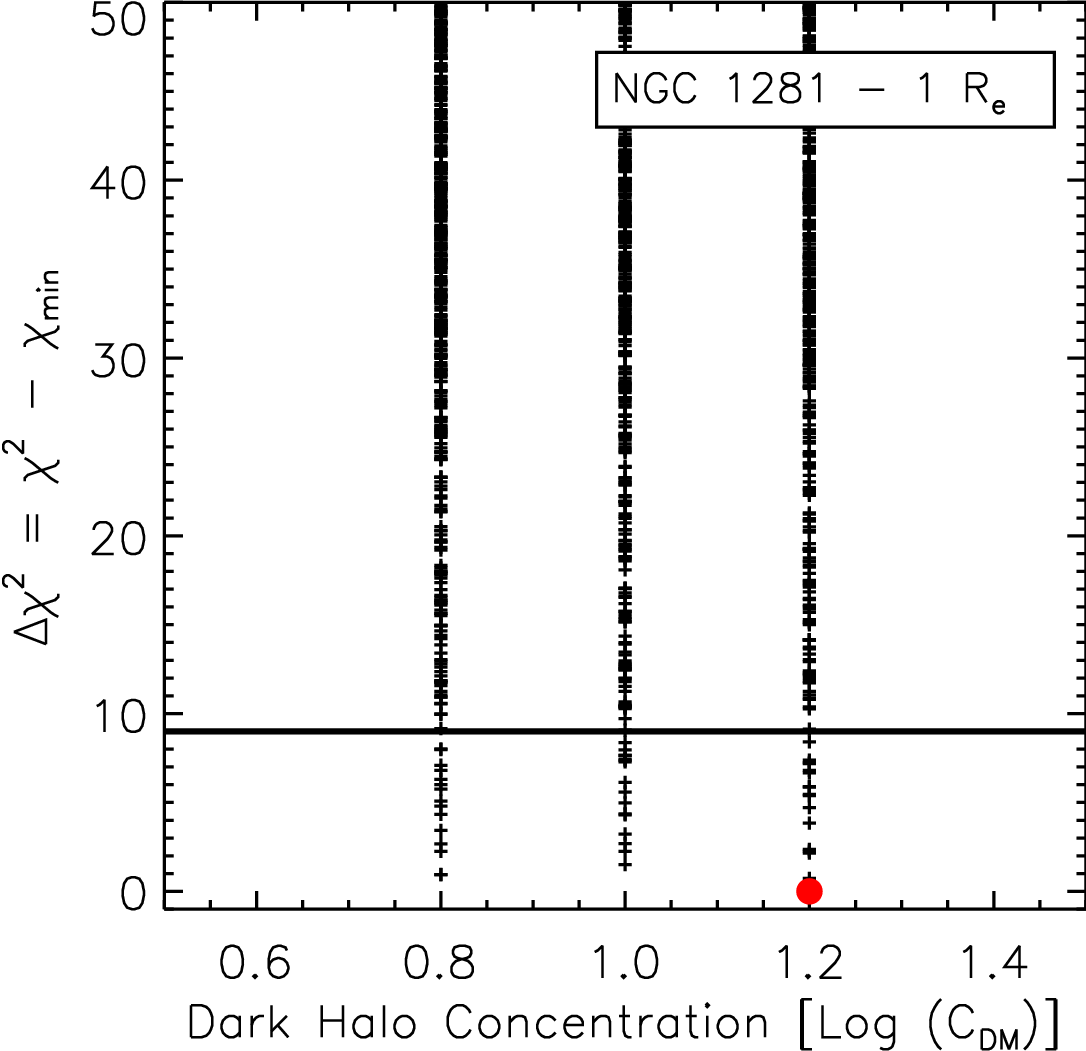}
		\includegraphics[width=.24\linewidth]{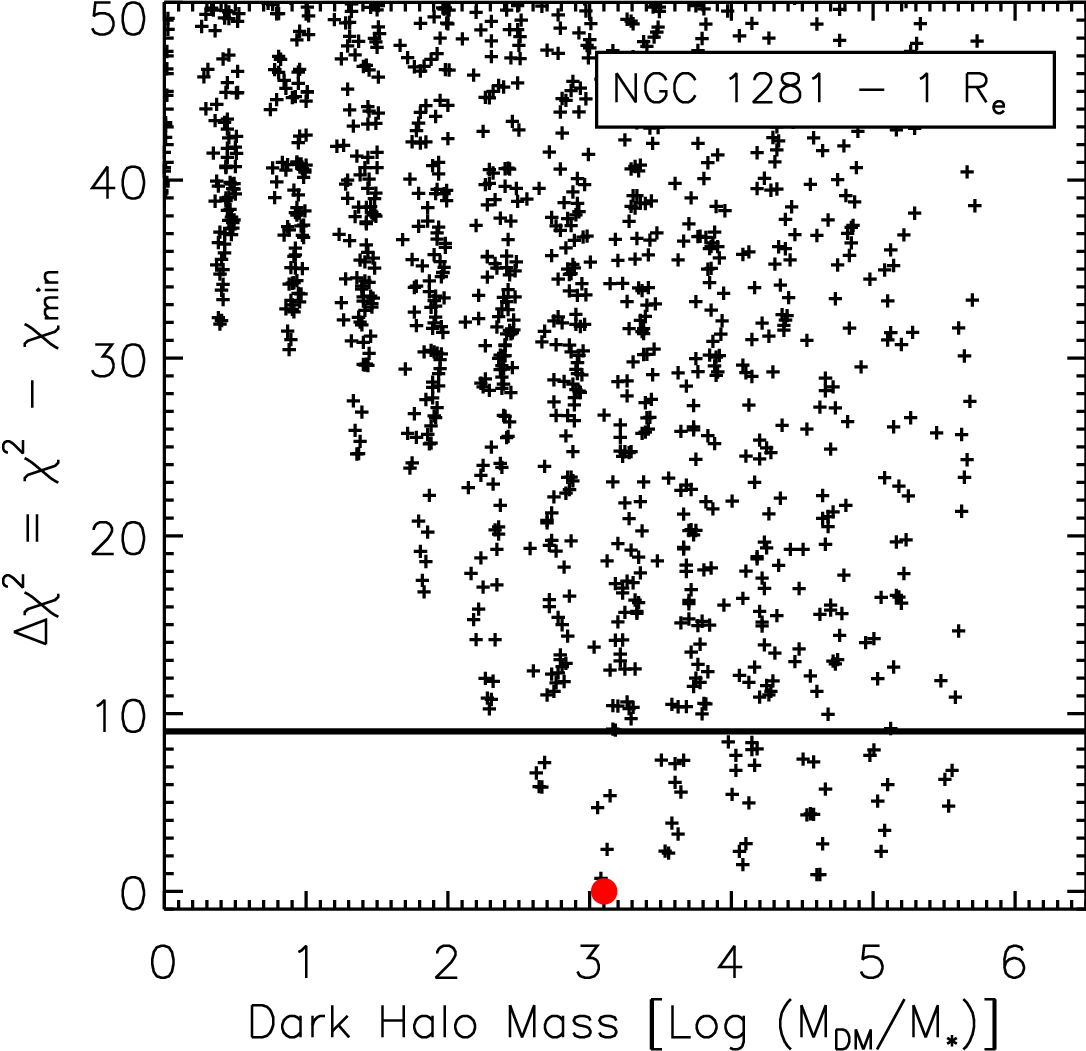}
		\includegraphics[width=.24\linewidth]{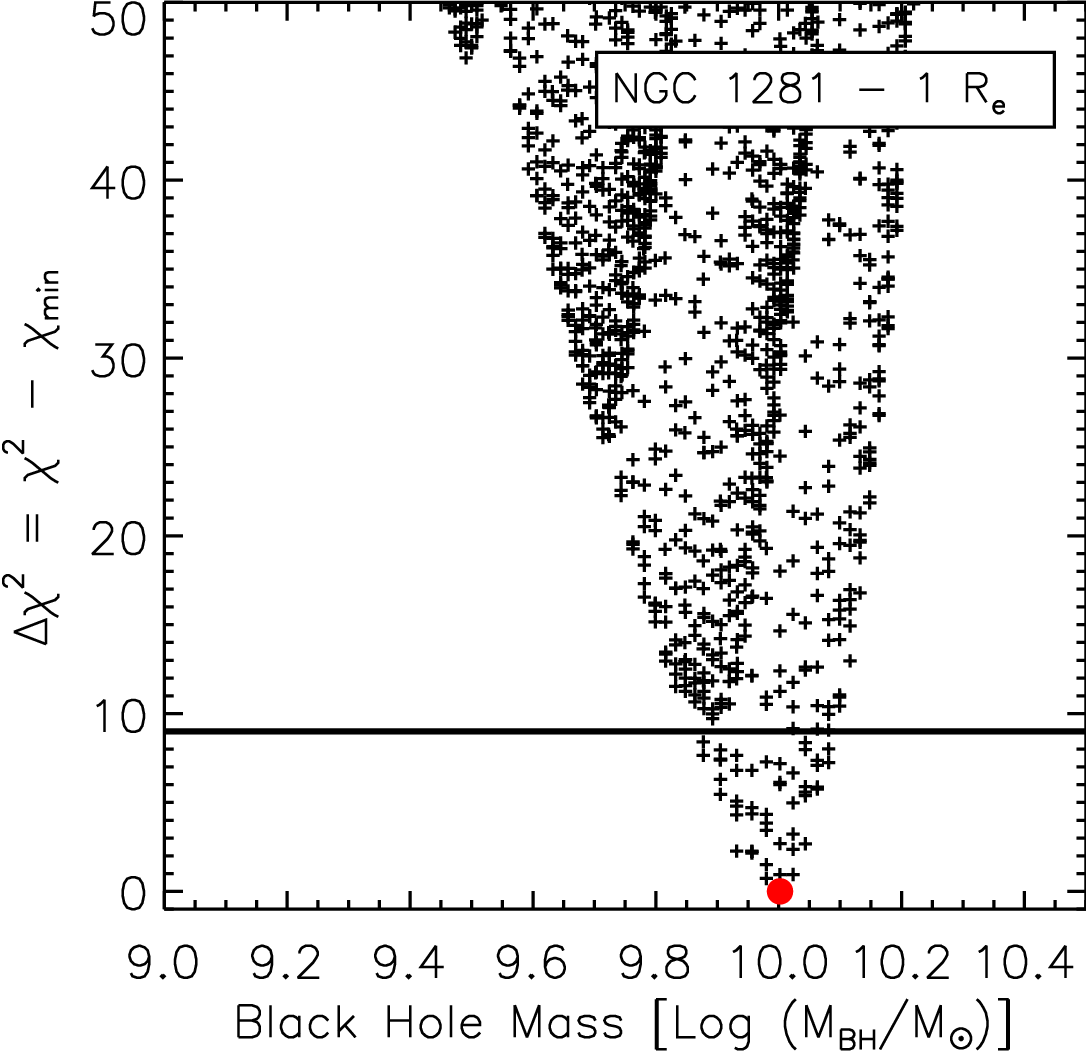}
		\includegraphics[width=.24\linewidth]{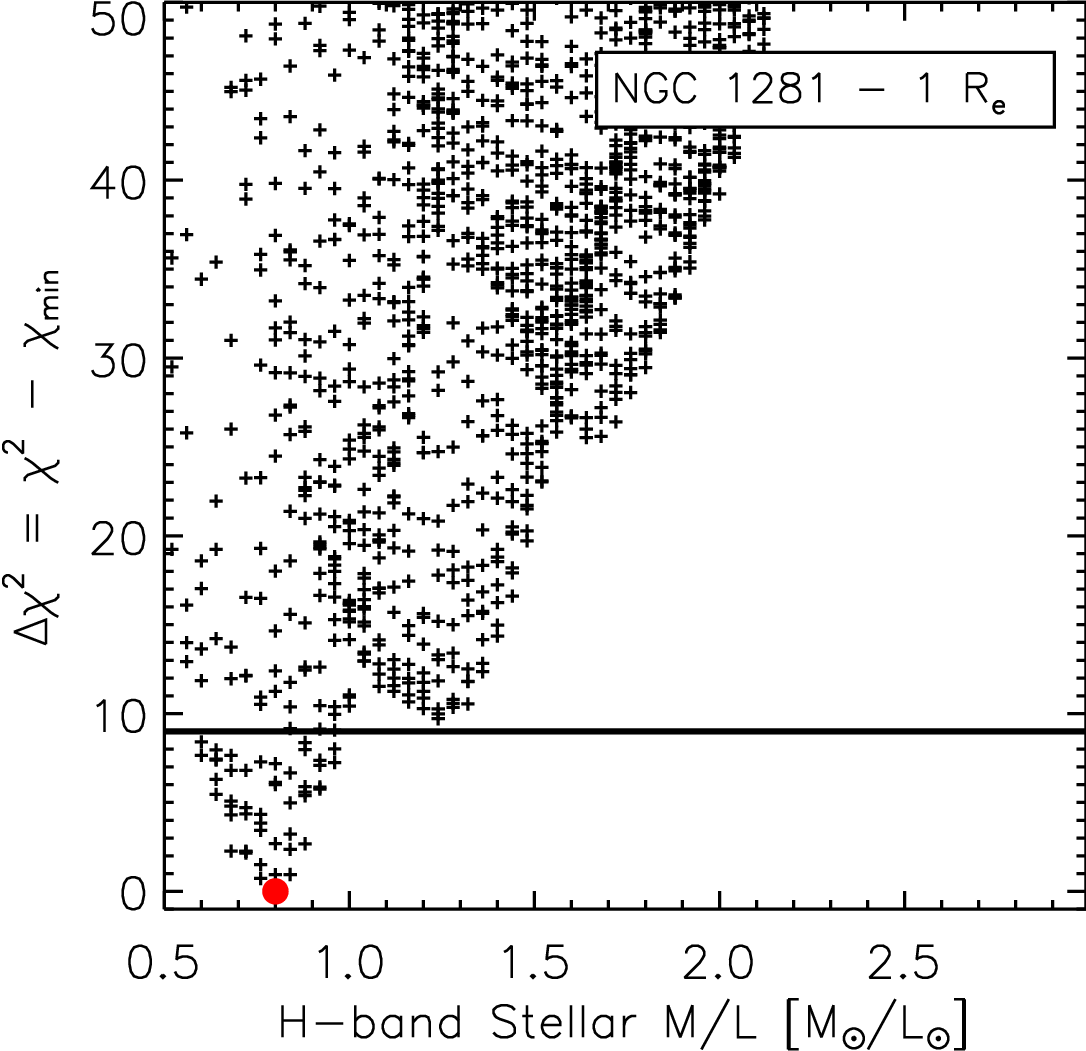}
\caption{Parameter constraints from our orbit-based dynamical models of NGC\,1281's wide-field IFU data within 5 (top panel) and 1 (bottom panel) \Reff, assuming a NFW halo. The columns illustrate the dark halo concentration log($c_{DM}$), dark halo mass to stellar mass ratio log($M_{DM}$/\Mstar), black hole mass log(\Mbh/$M_{\scriptstyle \odot}$) and stellar mass-to-light ratio in $H$-band. The horizontal line denotes a $\Delta \chi^2$ difference of 9, which corresponds to statistical 3$\sigma$ uncertainties for one degree of freedom, and the red dot corresponds to the overall best-fitting model parameter. The figures illustrate the constraints on the massive dark halo by means of the large scale stellar kinematics, which in turn result in an unreasonably low stellar M/L and over-massive SMBH.}
\label{fig:ngc1281_parameters}
\end{figure*}

The best-fitting dynamical model prefers an extremely massive SMBH, a massive dark halo and a very low stellar mass-to-light ratio. Fig. \hyperref[fig:ngc1281_model_kin]{\ref{fig:ngc1281_model_kin}} points at the outer velocity moments as the driver of these detections, and indeed model predictions beyond 1 \Reff\ show deviations from the observations if more typical values for the stellar mass-to-light ratio, dark halo and black hole are adopted, thus underlining the power of the wide-field IFU data here when compared to more traditional, shallow long-slit stellar kinematics. For instance, while both the best-fitting model with and without a halo can still fit most measurements within the 1$\sigma$ uncertainties, models without a halo systematically underpredict $\sigma$ and overpredict the absolute $h_3$ values in the outermost bins. On the other hand, both models fail to recover $\sigma$ within the central 3\arcsec, with velocity dispersion profiles that are too steep and which overshoot the 1$\sigma$ measurement errors. In the models without a halo, this can be traced back to the high stellar mass-to-light ratio. In order to optimally fit the outer LOSVD, the stellar mass contribution is maximised and overpredicts the stellar velocity dispersion profile in the very centre. Models with a halo can recover the observations better by ascribing much of the total mass to the dark halo, but - if parameterised by a spherical NFW profile - call for a significant amount of dark matter within the central arcseconds and equally well overestimates the central total mass content, despite driving down the stellar M/L by a factor of 4.\\

The constraints by the remote kinematic measurements are further exemplified in Fig. \hyperref[fig:ngc1281_parameters]{\ref{fig:ngc1281_parameters}}, where we show the estimates for the individual parameters in the fit of our fiducial models and test models which fit the data only within 1 \Reff. In our fiducial models, which fit the full LOSVD, both the black hole mass and stellar mass-to-light ratio are tightly constrained. As discussed above, this is a result of the massive dark halo which propagates inwards and drives down the stellar M/L, trying avoid an overestimation of the total mass content within the effective radius. The low stellar M/L, in turn, entails a massive SMBH for the recovery of the mass within the central 3\arcsec. Even so, the central velocity dispersion profile in the best-fitting NFW model exceeds the measurements and intuitively one would expect the models to infer a lower black hole mass in order to account for the mismatch. But, simply decreasing the black hole mass does not provide a better fit due to the nested interplay of the three individual mass components (see also Sec. \hyperref[sec:ngc1281_bh]{\ref{sec:ngc1281_bh}}), which are almost equally massive at 3\arcsec\ (Fig. \hyperref[fig:ngc1281_model_mass]{\ref{fig:ngc1281_model_mass}}). Usually, dark matter is negligible at sub-kpc scales and the black hole becomes degenerate with the stellar M/L only \citep[e.g.][]{2015ApJ...808..183W}. Decreasing the black hole would in this case, however, either imply a rise in the stellar M/L or the dark matter. The attempt to increase the stellar M/L faces a decrease in the dark matter, which in turn is tightly constrained by the remote kinematic measurements. Vice versa, the necessary level of increase in the dark matter within the central arcseconds would amount to a massive increase in the total dark content (under the assumption of a NFW profile), and thus fail to fit the outer LOSVD again.

Whereas the contours of the dark halo mass are well defined, the exact halo shape is not. Within the statistical uncertainties, the models can adopt any concentration within the range that is probed (see also Sec. \hyperref[sec:trunc]{\ref{sec:trunc}} for a discussion of the dark halo concentration). Moreover, the scale radius is 200\arcsec\ $\le \, r_s \, \le$ 3500\arcsec\ (i.e. $60$\,kpc $\le \, r_s \, \le$ 1000\,kpc) and hence promotes a constant density slope of the dark halo within the kinematic extent. Limiting the LOSVD to the measurements within one effective radius provides consistent results with the fiducial models, however, with larger confidence intervals for the dark halo.  It is also worth noting that the influence of the massive dark halo is already noticeable within the central 5\arcsec, where the sharp rise in the enclosed mass profile can only be explained by an additional dark component.\\

The orbital structure of our best-fitting NFW model is close to isotropic within the reach of the wide-field IFU data and becomes slightly tangentially anisotropic within 1\arcsec\, which, however, is not resolved by our data and merely an extrapolation by the models. The orbit distribution is also dominated by short-axis tube orbits, as is expected for an axisymmetric system, with box orbits being a significant contributor only in the innermost region, due to the perturbation of the axisymmetric potential close to the black hole \citep{2010MNRAS.401.1770V}.

%============================= section 4 =============================
\section{Discussion}
\label{sec:discussion}
%=====================================================================

According to our orbit based dynamical models of the wide-field IFU data, NGC\,1281 cannot be recovered without the presence of a dark halo (Fig. \hyperref[fig:ngc1281_model_kin]{\ref{fig:ngc1281_model_kin}} and \hyperref[fig:ngc1281_parameters]{\ref{fig:ngc1281_parameters}}). Models with a dark halo provide a good fit to the LOSVD and yield a total enclosed mass of 11.5 $\le$ log($M_{TOT}/M_{\scriptstyle \odot}$) $\le$ 11.8 within the kinematic extent. Given it's total luminosity and a stellar M/L of $\le$ 1 \MLsun\ (as indicated by our best-fitting dynamical model), this results in a dark matter contribution of 90 per cent to the total enclosed mass within 5 \Reff\ ($\simeq$ 7\,kpc). Adopting an optimistic stellar M/L of $\sim$ 2 \MLsun\ in \textit{H}-band would still imply a substantial halo to stellar mass ratio of 0.8. Based on these arguments, we can conclude that NGC\,1281 is embedded in a massive dark halo, even if we would choose to remain ignorant regarding the details of its dark halo profile, black hole mass and stellar M/L.\\

Despite being dominated by dark matter, the low stellar M/L and high black hole mass cast doubt on the veracity of the adopted dark halo profile. In the following subsections we will therefore explore the possibility of alternative stellar and dark halo parameterisations, while fitting the full LOSVD out to 5 \Reff, for which we summarise the modelling results in Table \hyperref[tab:all]{\ref{tab:all}}.

%---------------------------------------------------------------------
\subsection{Alternative Halo Parameterisations}
\label{sec:altern_halo}
%---------------------------------------------------------------------

%---------------------------------------------------------------------
\subsubsection{Cored-Logarithmic Halo and Mass-Follows-Light}
\label{sec:cl_halo}
%---------------------------------------------------------------------

\begin{table*}
	\caption{Best-fitting parameters and statistical 3$\sigma$ uncertainties for the respective set of models, with a spherical NFW halo (i), a cored-logarithmic halo (ii), mass-follows-light (iii), a spherical NFW halo including the presence of stellar M/L gradients (iv), a spherical NFW halo with a fixed stellar M/L according to our stellar population synthesis fits (v), and a spherical NFW halo with a fixed black hole mass according to the black hole mass - stellar velocity dispersion relation (vi). The columns indicate the difference in $\chi^2$ of the best-fitting model with respect to the best-fitting reference model (bold), the corresponding best-fitting M/L, black hole mass and dark matter content within the kinematic bounds of 5 \Reff\ or 7\,kpc. Note, that the number provided in row (iv) is the pivot stellar M/L for models with a spatially varying mass-to-light ratio of slope $\kappa = 0.1$.}

	\begin{center}
	\centerline{
	\begin{tabular}{ c | c  c  c  c }
		\hline
		& $\Delta \chi^2$ & \MLstar/\MLsun & log(\Mbh/$M_{\scriptstyle \odot}$) & log($M_{DM}$/$M_{\scriptstyle \odot}$)\\
		\hline
		\textbf{NFW} (i) & \textbf{0} & \textbf{0.7$^{+0.4}_{-0.1}$} & \textbf{10.0$^{+0.1}_{-0.2}$} & \textbf{11.7$^{+0.1}_{-0.2}$} \\
		& & & & \\
		Cored-Log (ii) & 14.7 & 1.0$^{+0.1}_{-0.1}$ & 10.0$^{+0.1}_{-0.1}$ & 11.8$^{+0.1}_{-0.4}$ \\
		& & & & \\
		Mass-Follows-Light (iii) & 114.1 & 2.7$^{+0.1}_{-0.1}$ & 8.1$^{+0.9}_{-8.1}$ & 0.0\\
		& & & & \\
		M/L Slope (iv) & -0.6 & 0.7$^{+0.6}_{-0.2}$ & 10.0$^{+0.0}_{-0.2}$ & 11.4$^{+0.2}_{-0.6}$ \\
		& & & & \\
		Stellar Populations M/L (v) & 35.6 & 1.7 & 9.7$^{+0.1}_{-0.1}$ & 11.2$^{+0.1}_{-0.2}$ \\
		& & & & \\
		Fixed BH (vi) & 80.8 & 2.2$^{+0.2}_{-0.2}$ & 8.9 & 11.0$^{+0.2}_{-0.4}$\\
		\hline
	\end{tabular}
	}
%	\vspace{2ex}
	\label{tab:all}
	\end{center}

\end{table*}

\begin{figure}
		\begin{center}
		\includegraphics[width=.45\textwidth]{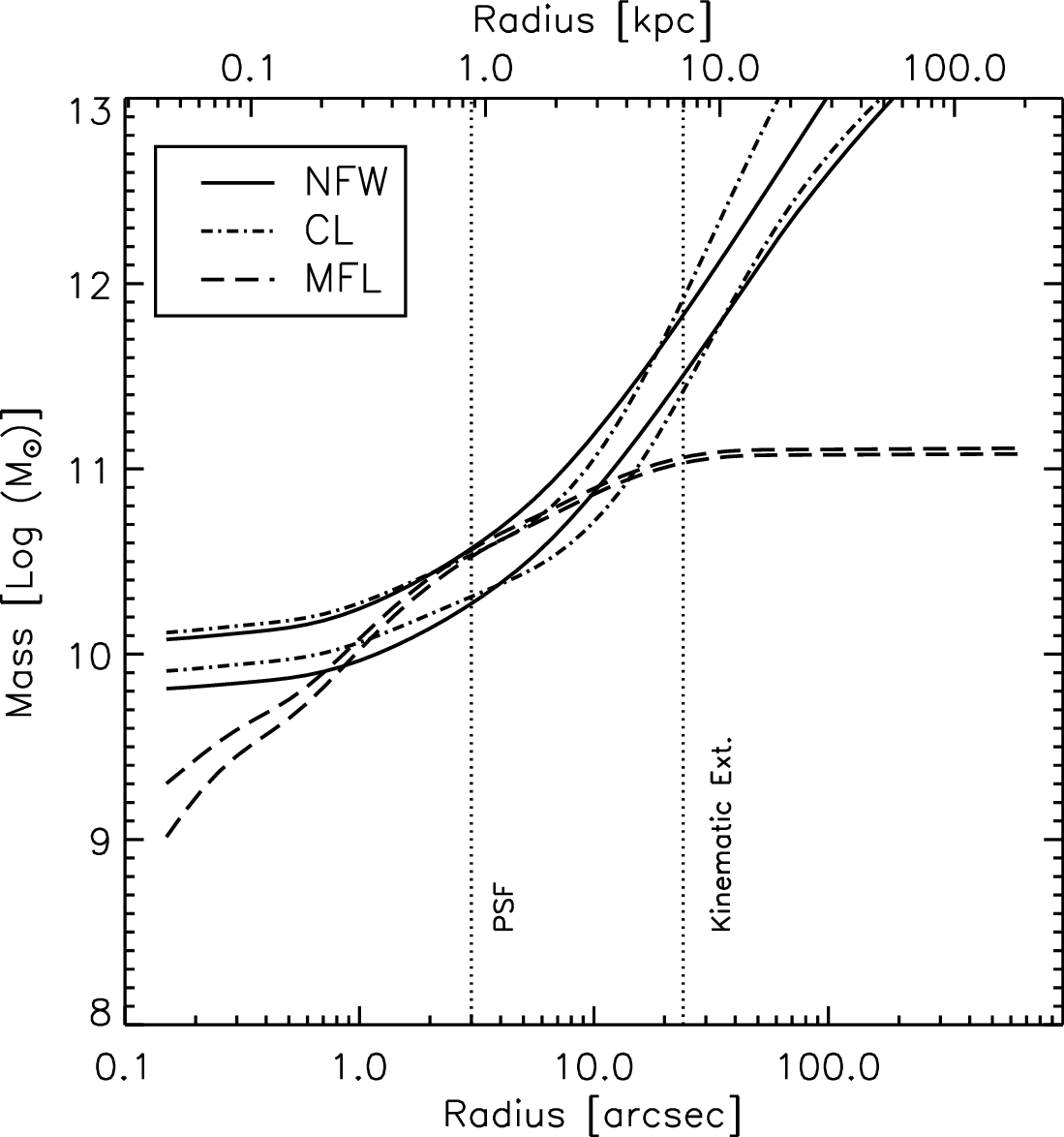}
		\end{center}
	\caption{Enclosed total mass as a function of radius for models with a NFW halo (solid), a cored-logarithmic halo (dash dot) and where mass-follows-light (long dash). The dotted vertical lines depict the resolution limit and the extent of the \ppak\ kinematics. The upper and lower line for each halo assumption corresponds to the upper and lower bound of the statistical 3$\sigma$ uncertainties. The cored-logarithmic and NFW halo parameterisations are in agreement, whereas the mass-follows-light models underestimate the total enclosed mass at the maximum kinematic extent.}
	\label{fig:ngc1281_enclosed}
\end{figure}

The dark halo mass in NGC\,1281 is based on the assumption that the dark matter density profile is well approximated by a spherically symmetric NFW profile. To assess the robustness of the results with respect to changes in the halo parameterisation, we have constructed models with a spherical logarithmic halo \citep{1987gady.book.....B}. The logarithmic halo density distribution is given by:
\begin{equation}
\rho(r) = \frac{v_c^2}{4 \pi G} \, \frac{(3 r_c^2 + r^2)}{(r_c^2 + r^2)^2}
\end{equation}
where $r_c$ is the core radius and $v_c$ the halo circular velocity, sampling parameter ranges of log($r_c$/kpc) $\in$ [1,2] and log($v_c$/\kms) $\in$ [1,4].
The cored-logarithmic halo possesses a flat central density core inside the core radius and gives rise to an asymptotic constant circular velocity at very large radii . In Fig. \hyperref[fig:ngc1281_enclosed]{\ref{fig:ngc1281_enclosed}} we plot the enclosed total mass as a function of radius for models with a NFW halo, a cored-logarithmic halo (CL) and, in addition, for models where mass-follows-light (MFL). The upper and lower line for each halo parameterisation corresponds to the upper and lower bound 3$\sigma$ statistical uncertainty in the respective set of models.

Within the uncertainties the CL halo has a core radius of 60\arcsec\ $\le \, r_c  \, \le$ 110\arcsec\ (i.e. $17$\,kpc $\le \, r_c \, \le 56$\,kpc) and a circular velocity of $1700$ \kms\ $\le \, v_c \, \le\ 5600$ \kms. Here again, the models prefer a constant slope for the dark matter density distribution within the kinematic bounds and yield a stellar mass-to-light ratio of 0.9 $\le$ \MLstar/\MLsun\ $\le$ 1.1 in \textit{H}-band. The exceptionally high circular velocity  $v_c$, on the other hand, reflects the massive dark halo of NGC\,1281. For comparison, orbit-based dynamical models of massive Coma ETGs as well as of individual elliptical and brightest cluster galaxies, such as M\,87, hint at more modest circular velocities of $100 \le v_c \le 800$ \kms\ \citep{1997ApJ...488..702R,2007MNRAS.382..657T,2011ApJ...729..129M}.

The total enclosed mass of the NFW and CL models are in excellent agreement, being about 11.4 $\le$ log($M_{TOT}/M_{\scriptstyle \odot}$) $\le$ 11.9 within 24\arcsec\ ($\simeq$ 5 \Reff\ or 7\,kpc), and the slight deviations are an effect of the coarse sampling strategy in the latter. The simple mass-follows-light models fail to keep pace and predict a total enclosed mass of 11.0 $\le$ log($M_{TOT}/M_{\scriptstyle \odot}$) $\le$ 11.1 within the same spatial extent. Moreover, the best-fitting MFL model has a mass-to-light ratio of 2.7 and deviates by $\Delta\chi^2 \sim 114$ from the best-fitting NFW model, as illustrated in Fig. \hyperref[fig:ngc1281_model_kin]{\ref{fig:ngc1281_model_kin}}.\\

%---------------------------------------------------------------------
\subsubsection{Spatially Varying Stellar M/L}
\label{sec:ngc1281_ml}
%---------------------------------------------------------------------

The most restrictive condition in our dynamical models is the assumption of a constant \MLstar. The stellar mass-to-light ratio in NGC\,1281 is 0.6 $\le$ \MLstar/\MLsun\ $\le$ 1.1 and much lower than expected from an old stellar population ($\ge$ 10 Gyr) of a canonical Kroupa (\MLstar/\MLsun\ = 1.3) or Salpeter IMF (\MLstar/\MLsun\ = 1.7, see Sec. \hyperref[sec:ngc1281_imf]{\ref{sec:ngc1281_imf}}). A radial variation of the mass-to-light ratio, however, could mitigate the effect of missing mass in the remote regions and lower the predictions for additional dark mass while increasing \MLstar.

To this end we constructed dynamical models with a \MLstar\ gradient, covering single slopes of $0 < |\kappa| \le 0.5$. The models also include a SMBH and a spherical NFW halo and the sampling in parameters space has not changed with respect to our fiducial models in Sec. \hyperref[sec:ngc1281_schwarzschild]{\ref{sec:ngc1281_schwarzschild}}. Here, we explicitly exclude models with $\kappa = 0$ from this set, as they already correspond to our fiducial models with a NFW halo and a constant \MLstar. The functional form of the stellar mass-to-light ratio is then given by log($\Upsilon_{\alpha}$) = log($\Upsilon_{\alpha, 0}$) + $\kappa$ $\times$ log ($r/r_0$), with $r$ being the radius in arcseconds and $\Upsilon_{\alpha, 0}$ the pivot stellar M/L at $r_0$ = 0.1\arcsec. We found that models with slopes of $\kappa$ = 0.1$\pm$0.2 can recover the observations very well. From a $\chi^2$ point of view, the best-fitting model with a slope provides an almost equally good fit as the best-fitting fiducial model, with $\chi^2_{NFW} - \chi^2_{SLOPE} = -0.6$ (Tab. \hyperref[tab:all]{\ref{tab:all}}). In spite of the increase in the degrees of freedom, with two additional model fitting parameters with respect to the fiducial NFW models, the only slight decrease in $\chi^2$ of these models is simply related to the coarse sampling strategy in $\kappa$ and the fact that we haven't yet found the absolute minimum in its $\chi^2$ distribution.

In general, the slopes are positive and driven by the need of a substantial amount of dark mass. The radial increase of the stellar mass-to-light ratio, however, cannot account for a significant fraction of the total dynamical mass due to the rapid decline of the stellar surface density of NGC\,1281. Hence the lower bound predictions for the dark halo virial mass decrease only by 0.5 dex, with the best-fitting model now located at log($M_{DM}$/\Mstar) = 4.1, log(\Mbh/$M_{\scriptstyle \odot}$) = 10.0, $\Upsilon_{\alpha, 0}$ = 0.7 \MLsun\ and $\kappa$ = 0.1.

\begin{figure}
		\begin{center}
		\includegraphics[width=.45\textwidth]{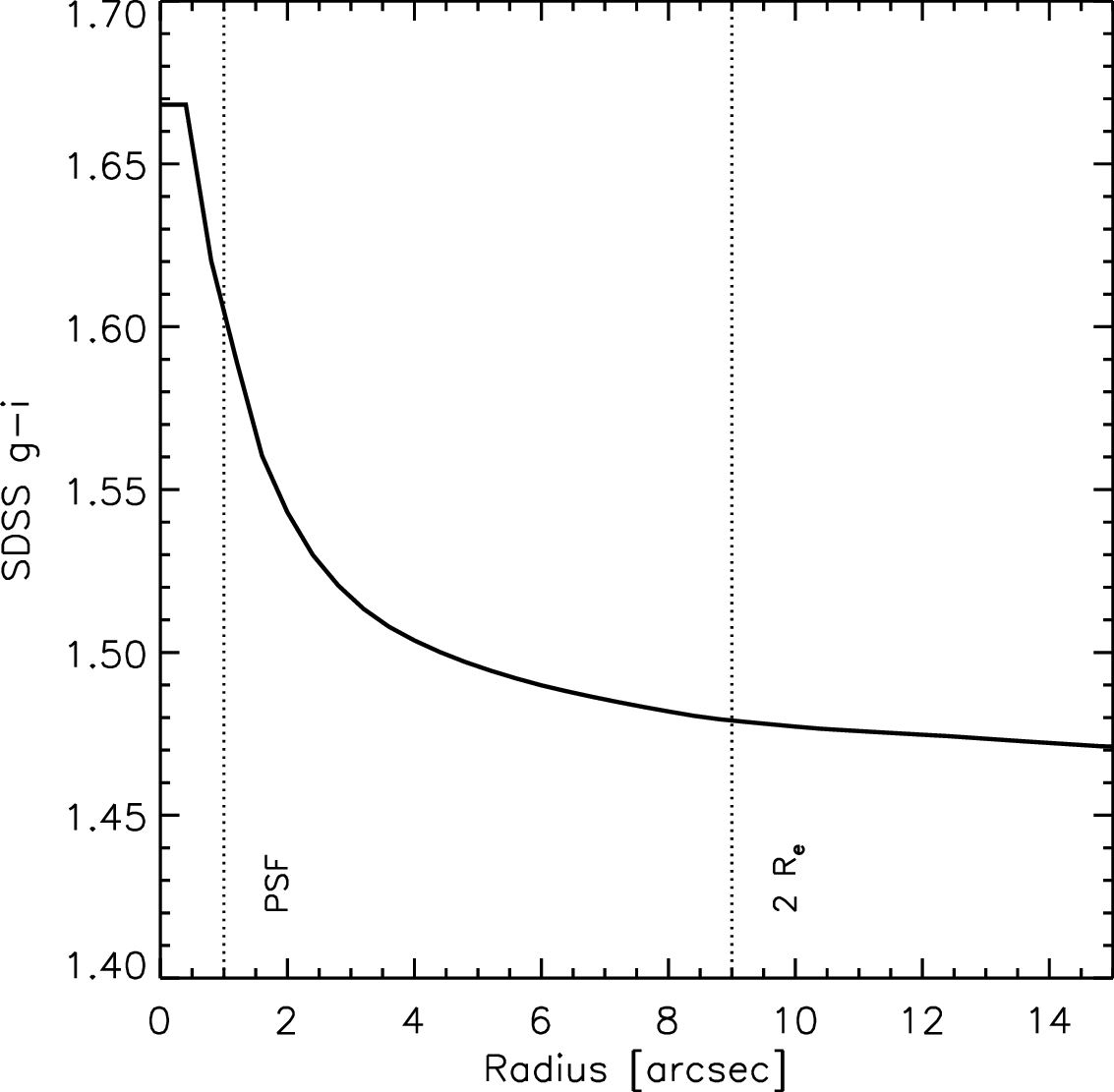}
		\end{center}
	\caption{SDSS g-i colour profile of NGC\,1281. The first line is the mean PSF FWHM of the two particular SDSS bands and the second line depicts two effective radii as measured from the circular aperture of the high-resolution \textit{HST} imaging. Within this range, the colour decreases only by $\sim$ 0.1 mag and suggests no significant stellar M/L gradient.}
	\label{fig:ngc1281_colour}
\end{figure}

Moreover, a positive slope of 0.1 - 0.3 induces a change in the stellar mass-to-light ratio from 0.7 \MLsun\ in the centre to 1.2 \MLsun\ $-$ 3.6 \MLsun\ in the outermost regions. The existence of such gradients should be imprinted in the colour profiles of galaxies \citep{2001ApJ...550..212B}, attributable to strong variations in their stellar population properties, but is not evident in the $g - i$ colour profile of NGC\,1281 based on archival SDSS imaging (Fig. \hyperref[fig:ngc1281_colour]{\ref{fig:ngc1281_colour}}). A positive slope would also be at odds with the general trend of colour (and hence negative M/L) gradients in ETGs \citep{2010MNRAS.407..144T,2011MNRAS.418.1557T} in general and in NGC\,1281 in particular (see Sec. \hyperref[sec:ngc1281_imf]{\ref{sec:ngc1281_imf}}). The profile shows only a very gentle drop within 2 effective radii of about 0.1 mag, indicative of a negative gradient with an expected variation in the stellar M/L of 0.2 \citep{2009MNRAS.400.1181Z}, which shows that the use of a constant \MLstar\ was a reasonable choice.\\

%---------------------------------------------------------------------
\subsubsection{The Stellar Populations M/L}
\label{sec:ngc1281_imf}
%---------------------------------------------------------------------

NGC\,1281 is a compact, high velocity dispersion galaxy and bears resemblance to the compact objects at redshift $z = 2$, which are thought to be the progenitors of the inner parts of today's massive elliptical galaxies. It is also very similar to NGC\,1277, another high-dispersion galaxy which hosts an over-massive SMBH for its modest dimensions \citep{2012Natur.491..729V,2015MNRAS.452.1792Y,WBGY}. The IMF of these objects could be significantly different from the IMF observed in the local volume. In fact, the universality of the IMF has been questioned recently, based on results from gravitational lensing \citep[e.g.][]{2010ApJ...709.1195T,2010ApJ...721L.163A}, stellar dynamics \citep[e.g.][]{2012Natur.484..485C,2012MNRAS.422L..33D,2013MNRAS.432.2496D,2013ApJ...765....8T} and SPS fits to gravity sensitive features in the integrated spectra of ETGs \citep[e.g.][]{2010Natur.468..940V,2012ApJ...760...71C,2014MNRAS.438.1483S,2015MNRAS.447.1033M,2015MNRAS.451.1081M}. These results suggest a bottom heavy IMF with increasing velocity dispersion which should lead to stellar mass-to-light ratios larger than those predicted by a Salpeter IMF, assuming that the shape of the IMF in these objects is similar to what is observed in the Galaxy. However, we have shown that models with high \MLstar\ can be excluded as they are not able to recover the features of the LOSVD and miss to replicate the high dynamical mass of NGC\,1281, if the dark halo is parameterised by a NFW or CL profile.
 
\begin{figure}
		\begin{center}
		\includegraphics[width=.45\textwidth]{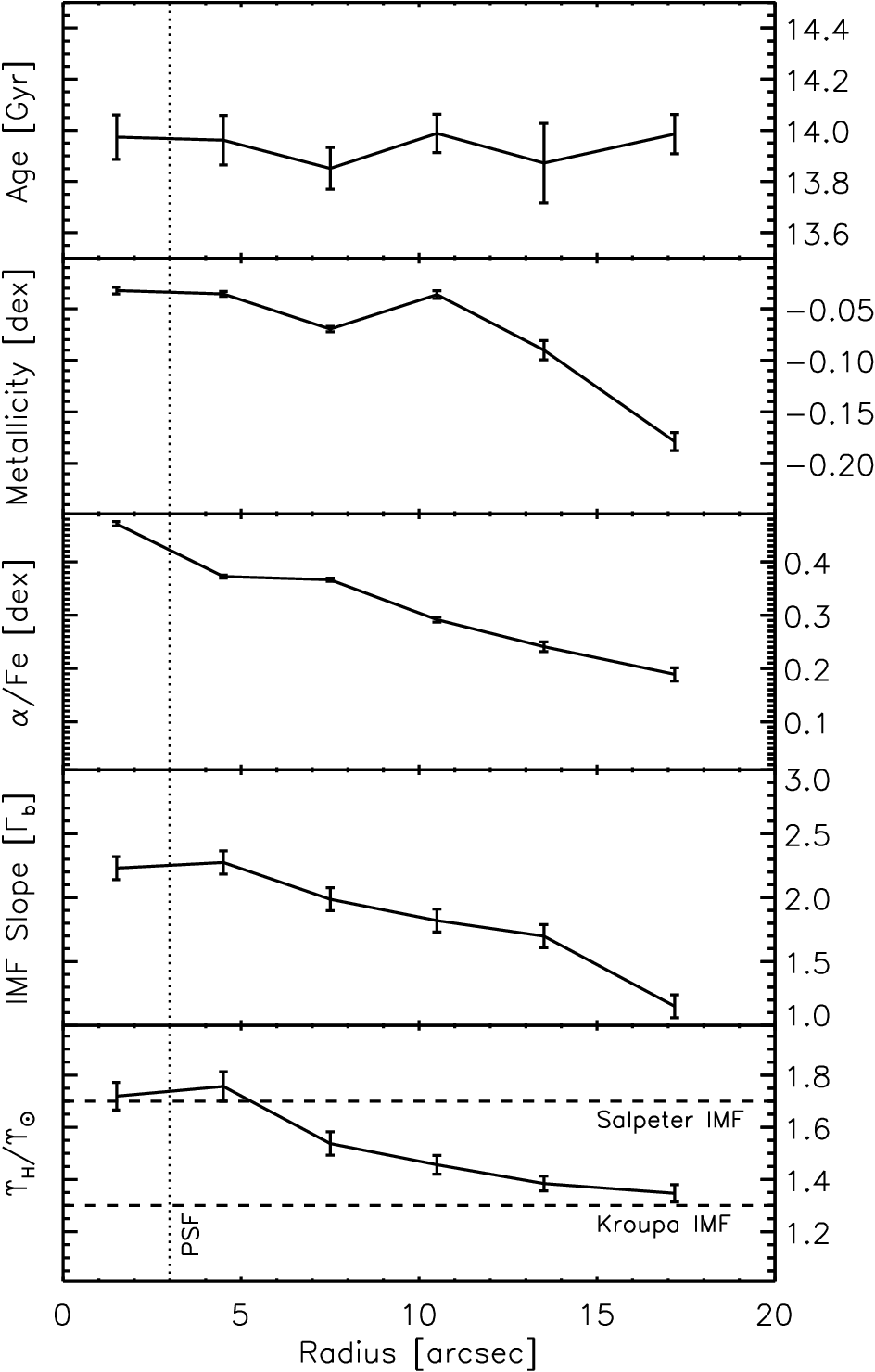}
		\end{center}
	\caption{Stellar population properties of NGC\,1281. The top three panels represent the age, metallicity and $\alpha$-abundance as a function of radius. The fourth panel depicts the IMF slope. Under the assumption of a bimodal IMF, these stellar population properties can be turned into a stellar mass-to-light ratio (bottom panel). Constrained by the uniformly old stellar population, the resulting stellar mass-to-light ratios range between 1.7 \MLsun\ (centre) and 1.3 \MLsun\ (outskirts) in \textit{H}-band, and can not be reconciled with the stellar M/L of 0.6 $\le$ \MLstar/\MLsun\ $\le$ 1.1 in the NFW models.}
	\label{fig:ngc1281_stellar_pops}
\end{figure}

To obtain an independent measurement of the stellar M/L and to asses the reliability of the stellar M/L in our dynamical set of models with a NFW profile, we carried out a line-strength analysis of NGC\,1281 based on our \ppak\ observations. We made use of the extended MILES stellar population models \citep{2010MNRAS.404.1639V,2012MNRAS.424..157V}, covering a range from -2.32 dex to +0.22 dex in metallicity and from 0.06 Gyr to 17 Gyr in age. We allowed the IMF slope to vary, from $\Gamma_b$ = 0.3 to $\Gamma_b$ = 3.3, assuming a bimodal (low-mass tapered) IMF parameterisation. The analysis suggests a very old ($\sim$14 Gyr) stellar population throughout the galaxy, which is consistent with the star formation history derived in \cite{2015ApJ...808...79F} based on SDSS spectroscopic data (Fig. \hyperref[fig:ngc1281_stellar_pops]{\ref{fig:ngc1281_stellar_pops}}). The metallicity decreases from solar in the centre to -0.17 dex at 3 \Reff, and the IMF slope, parameterised following \cite{1996ApJS..106..307V} and \cite{2013MNRAS.433.3017L}, varies from $\Gamma_\mathrm{b}$ = 2.3 in the central regions to $\Gamma_\mathrm{b}$ = 1.3 in the outskirts. Given these best-fitting stellar population parameters - and under the assumption of a bimodal IMF - the expected radial variation in the $H$-band stellar M/L is $\Delta \Upsilon_\star = 0.4$ \MLsun, where the extra variation of 0.2 with respect to the variation indicated by the colour profile (Fig. \hyperref[fig:ngc1281_colour]{\ref{fig:ngc1281_colour}}) is related to the fact that we allowed the IMF to vary. Such a radial decrease in the stellar M/L is consistent with the analysis of the compact, elliptical galaxy NGC\,1277 \citep{2015MNRAS.451.1081M}, where a mild but negative radial gradient was also found. Note also that a positive gradient (as discussed in Sec. \hyperref[sec:ngc1281_ml]{\ref{sec:ngc1281_ml}}) is not compatible with the stellar population analysis of NGC\,1281 in particular and with ETGs in general, since metallicity and IMF slope track each other \citep{2015MNRAS.447.1033M} and which in both tend to decrease radially. Although different IMF parameterisations could lead to different M/L values (while fitting equally well the strength of gravity sensitive features), a M/L as low as inferred from the dynamical analysis (\MLstar/\MLsun = 0.7) is only possible under the assumption of a very young ($\sim$ 5 Gyr) stellar population. However, the presence of such a young stellar component is ruled out by the line-strength analysis and the SDSS $g-i$ colour profile.

Adopting a stellar mass-to-light ratio of 1.7 \MLsun, as indicated by the stellar populations, is capable to decrease the halo to stellar mass ratio to log($M_{DM}$/\Mstar) = 2.9. Yet, fixing \MLstar\ provides a worse fit to the kinematics than our overall best-fitting dynamical model (Fig. \hyperref[fig:ngc1281_model_kin]{\ref{fig:ngc1281_model_kin}}). Moreover, this model deviates by $\Delta\chi^2 \ge 30$ from our best-fitting model, and it remains elusive why one should prefer a model which is excluded by a significance of more than 5$\sigma$.

%---------------------------------------------------------------------
\subsubsection{The Black Hole Mass and \Mbh-$\sigma$ Relation}
\label{sec:ngc1281_bh}
%---------------------------------------------------------------------

The models predict the presence of a black hole of log(\Mbh/$M_{\scriptstyle \odot}$) = 10.0$^{+0.1}_{-0.2}$, but we strongly emphasise that this measurement is solely driven by the constraints for the dark halo and hence the stellar mass to light ratio. This chain of reactions starts with the necessity to recover the high total enclosed mass with a steeply rising NFW halo, which already dominates the total mass budget beyond 1 \Reff. The massive dark halo drives down the stellar M/L by about a factor of 2-3, when compared to our stellar population analysis, to relieve the tension for the dynamical mass within the effective radius. This, in turn, yields a tension between the low stellar mass and the total dynamical mass in the very centre, which gives rise to the unrealistic SMBH.

With the strong demand of an additional mass component, which is already noticeable from the fits within the effective radius (Fig. \hyperref[fig:ngc1281_parameters]{\ref{fig:ngc1281_parameters}}), the dark halo consequently joins the degeneracy in the centre between both the stellar and black hole mass. In this context, the mismatch between central velocity dispersion measurements and those that are recovered by our best-fitting dynamical models (Fig. \hyperref[fig:ngc1281_model_kin]{\ref{fig:ngc1281_model_kin}}) can simply be regarded as the vexing challenge to recover the total mass content and to provide sensible parameter estimates at the same time, when parameterising the dark halo with a NFW profile. As pointed out in Sec. \hyperref[sec:ngc1281_schwarzschild]{\ref{sec:ngc1281_schwarzschild}}, this threefold degeneracy prevents the models from adopting lower black hole masses by increasing the stellar and/or dark matter content, since this immediately translates into worse fits to the LOSVD beyond the central few arcseconds.\\

Even in our fiducial models, the black hole sphere of influence is only $R_{SOI} = 2.2$\arcsec\ and hence not resolved by our \ppak\ observations with an intrinsic fiber diameter of 2.7\arcsec\ and a PSF FWHM of $\sim$ 3\arcsec. Nevertheless, to investigate the impact of the central mass distribution, we have fixed the black hole mass according to the predictions of the \Mbh\ - $\sigma$ relation, adopting the central stellar velocity dispersion of 268 \kms. In this set of models, the black hole mass is assumed to be of the order of log(\Mbh/$M_{\scriptstyle \odot}) = 9.00$ \citep{2013ApJ...764..184M}. As expected, the lower black hole mass increases the stellar M/L to \MLstar\ = 2.2 \MLsun, to recover the central dynamical mass. Consequently, the dark halo contribution to the total mass profile decreases to log($M_{DM}$/\Mstar) = 3.0. However, the stellar M/L is now more massive than inferred from its stellar populations (Sec. \hyperref[sec:ngc1281_imf]{\ref{sec:ngc1281_imf}}). The models also fail to recover the details of the LOSVD, similar to what has been shown for MFL models in Fig. \hyperref[fig:ngc1281_model_kin]{\ref{fig:ngc1281_model_kin}}, and still overestimate the central dispersion peak.

\begin{figure}
		\begin{center}
		\includegraphics[width=.45\textwidth]{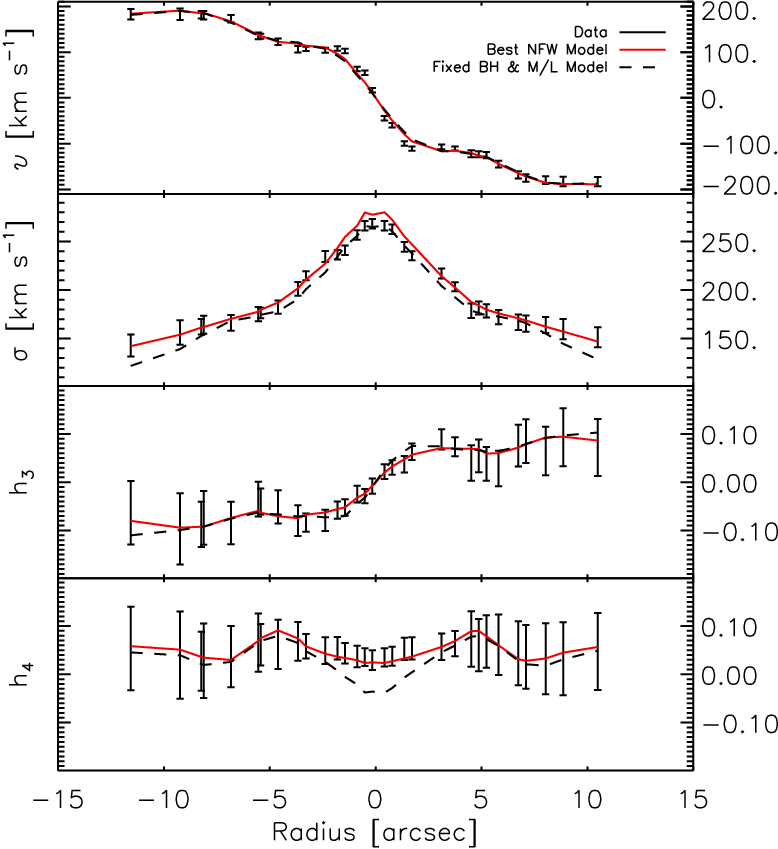}
		\end{center}
	\caption{LOSVD of NGC\,1281, for the Voronoi bin centroids which fall within a 1\arcsec wide strip along the major axis. The red and dashed line correspond to the recovered LOSVD of the best-fitting NFW model and a model with a fixed black hole mass of log(\Mbh/\Msun) = 9.1, a stellar M/L of 1.8 \MLsun\ and a dark halo of log($M_{DM}$/\Mstar) = 2.5. If fixed to parameters that are promoted by the black hole scaling relations and its stellar populations, the velocity dispersion peak can be recovered well, but the models fail to fit the outer velocity moments which stresses the need for a massive dark halo.}
	\label{fig:ngc1281_bh_ml}
\end{figure}

Interestingly, models with a moderate black hole mass and stellar mass-to-light ratio, which are preferred in light of the black hole scaling relations and our stellar population analysis, can match the central velocity dispersion much better (Fig. \hyperref[fig:ngc1281_bh_ml]{\ref{fig:ngc1281_bh_ml}}), but miserably fail to fit the outer velocity dispersions for low halo masses of log($M_{DM}$/\Mstar) $\le$ 2.5. This, once again, underscores that a massive, spherical NFW halo and a more reasonable black hole mass and stellar M/L are mutually exclusive in NGC\,1281. Only by excluding the seeing limited measurements, can we try to break the linkage between the innermost (black hole) and outermost (dark matter) mass component. And indeed, the stellar mass-to-light ratio and black hole mass start approaching more familiar territories, with parameter estimates of 0.6 $\le$ \MLstar/\MLsun\ $\le$ 2 and  9.3 $\le$ log(\Mbh/$M_{\scriptstyle \odot}$) $\le$ 10.2 within the statistical 3$\sigma$ uncertainties, when the central arcseconds are excluded from the fits.

It is also worth noting that the formidable reproduction of the central LOSVD with a black hole mass of log(\Mbh/$M_{\scriptstyle \odot}) = 9.1$, rather than log(\Mbh/$M_{\scriptstyle \odot}) = 10.0$, would still fit into a picture where the SMBHs in these compact ellipticals are over-massive with respect to their bulge luminosity or bulge mass, possibly due to the lack of structural evolution since z $\sim$ 2 \citep{2015ApJ...808...79F}.

%---------------------------------------------------------------------
\subsection{Conclusions, Caveats and Considerations}
\label{sec:caveats}
%---------------------------------------------------------------------

We obtain predictions for the dark halo to stellar mass ratio of log($M_{DM}$/$\Mstar$) = 3.6$^{+1.5}_{-0.5}$. The dark halo is not only one order of magnitude larger than the stellar mass within a comparatively small spatial extent of 5 \Reff\ ($\simeq$ 7\,kpc), but also more than one order of magnitude larger than what is expected if the mass-concentration relation \citep[e.g.][]{2008MNRAS.391.1940M} and stellar-to-halo mass relation \citep[e.g.][]{2010MNRAS.407....2D} holds. The effective radius in NGC\,1281 encompasses a dark matter fraction of $f_{DM}$ = 0.47$^{+0.13}_{-0.23}$, and as such, is 3 times higher than inferred from the population of regular ETGs in the ATLAS$^{3D}$ sample \citep{2013MNRAS.432.1709C} for galaxies with stellar masses in the range of $\sim$ log(\Mstar) =11. It is, however, in line with the dark matter fraction of BCGs such as NGC\,4889 \citep{2007MNRAS.382..657T} and NGC\,3842 \citep{2012ApJ...756..179M} and the mean dark matter fraction of massive lens galaxies within the SLACS sample \citep{2007ApJ...667..176G}, even if the latter approach commonly measures the dark matter content within a projected cylinder rather than within a sphere of radius \Reff\ and generally leads to higher dark matter fractions when compared to the dynamical mass estimate presented here. Furthermore, an extrapolation of the dark halo profile would yield a cluster sized halo, which would have to be embedded in the Perseus cluster halo itself. In the context of $\Lambda$CDM cosmology, though, this is very unlikely and in contrast to what has been observed so far based on results of weak gravitational lensing \citep{2006MNRAS.368..715M}, satellite kinematics \citep{2011MNRAS.410..210M} and halo abundance matching \citep{2010ApJ...710..903M}. Taking into account the prediction of an unreasonably low stellar M/L as well as the presence of an over-massive supermassive black hole, we conclude that a spherical NFW profile cannot be a fair assumption of the dark halo profile of this galaxy. The results for the NFW halo in our dynamical models can only be considered as an attempt to recover the high total enclosed dynamical mass ($M_{TOT}$) - which is not possible for simple MFL models, models with a spatial variation in the stellar M/L, models with a fixed M/L according to our spectral synthesis fits and models with a fixed black hole mass according to the \Mbh-$\sigma$ relation - without an interpretation of the physical consequences for the values of choice. This is also illustrated by the preference of a constant dark matter density slope - in both the NFW and CL case - since the models try to maximise the dark matter contribution within the observational extent. Moreover, the necessity of a rapidly rising dark halo contribution leads to a tension between the stellar and dark mass in the central parts, and this tension can only be lightened by significantly decreasing the stellar mass-to-light ratio to unrealistically low values (Sec. \hyperref[sec:ngc1281_imf]{\ref{sec:ngc1281_imf}}), which in turn calls for an unreasonably massive SMBH (Sec. \hyperref[sec:ngc1281_bh]{\ref{sec:ngc1281_bh}}).\\

%---------------------------------------------------------------------
\subsubsection{Concentrated, Contracted or Stripped}
\label{sec:trunc}
%---------------------------------------------------------------------

The presence of a constant inner density slope is not surprising, as IFU observations commonly fall short in providing kinematic data that reaches out far enough to probe the regions where the dark halo density slope turns over. In principle, the constant density slope could also be consistent with a NFW profile where the outer halo was stripped by the Perseus cluster environment. As the halo gets stripped, the collisionless dark matter particles could relax and the halo would puff up in the shallower gravitational potential well, similar to what has been predicted for the collisionless stellar components of dwarf NFW halos during the tidal disruption in MW like potentials \citep{2006ApJ...650L..33P}. The relaxation and puff up of the dark matter halo would shift the turnover radius even further away and out of reach for our limited FOV. But, even if the dark halo is truncated due to stripping, this does not solve the discrepancy between the halo mass of NGC\,1281 and the halo masses that are predicted within the $\Lambda$CDM framework and assigned to galaxies via the abundance matching mechanism. The dark halo of NGC\,1281 within 24\arcsec\ is still over-massive, and the relaxation due to stripping would imply that the unstripped halo was even more concentrated and massive before within the same spatial extent.\\

The exploration of the concentration parameter in our dynamical models is motivated by numerical $N$-body simulations of structure formation within $\Lambda$CDM cosmology \citep{2008MNRAS.391.1940M,2014MNRAS.441.3359D}. Our upper and lower bound of the concentration probes a scatter of $\sim$ 0.2 dex in log($c_{DM}$) around the best-fitting $c_{DM} - M_{200}$ relation, which at the same time encloses the 97.7th percentile of the distribution \citep{2008MNRAS.391.1940M}, assuming that the stellar-to-halo mass relation holds \citep{2010ApJ...710..903M}. Whereas X-ray observations indicate agreement between the concentrations that are found for galaxy clusters and those that are predicted by simulations \citep[e.g.][]{2005A&A...435....1P,2006ApJ...640..691V,2007ApJ...664..123B,2007MNRAS.379..209S}, individual lensing studies hint at slightly higher concentrations \citep{2003A&A...403...11G,2005ApJ...621...53B,2008ApJ...685L...9B,2012MNRAS.424..104L}. An increase of the upper bound of possible halo concentrations in our fiducial NFW models (see Sec. \hyperref[sec:ngc1281_schwarzschild]{\ref{sec:ngc1281_schwarzschild}}) to log($c_{DM}$) = 1.4, however, cannot conciliate our dynamical parameter estimates with our measurements from NGC\,1281's stellar populations and the expectations from numerical simulations. The lower bound of the dark halo to stellar mass ratio decreases to log($M_{DM}$/$\Mstar$) = 2.6, accompanied by a marginal increase in the upper bound of the stellar M/L to 1.2 \MLstar. With no sign of convergence for the halo concentration, so no evidence for a more concentrated spherical NFW halo, we refrain from probing even higher values.\\

\begin{figure}
		\centering
		\includegraphics[width=.49\linewidth]{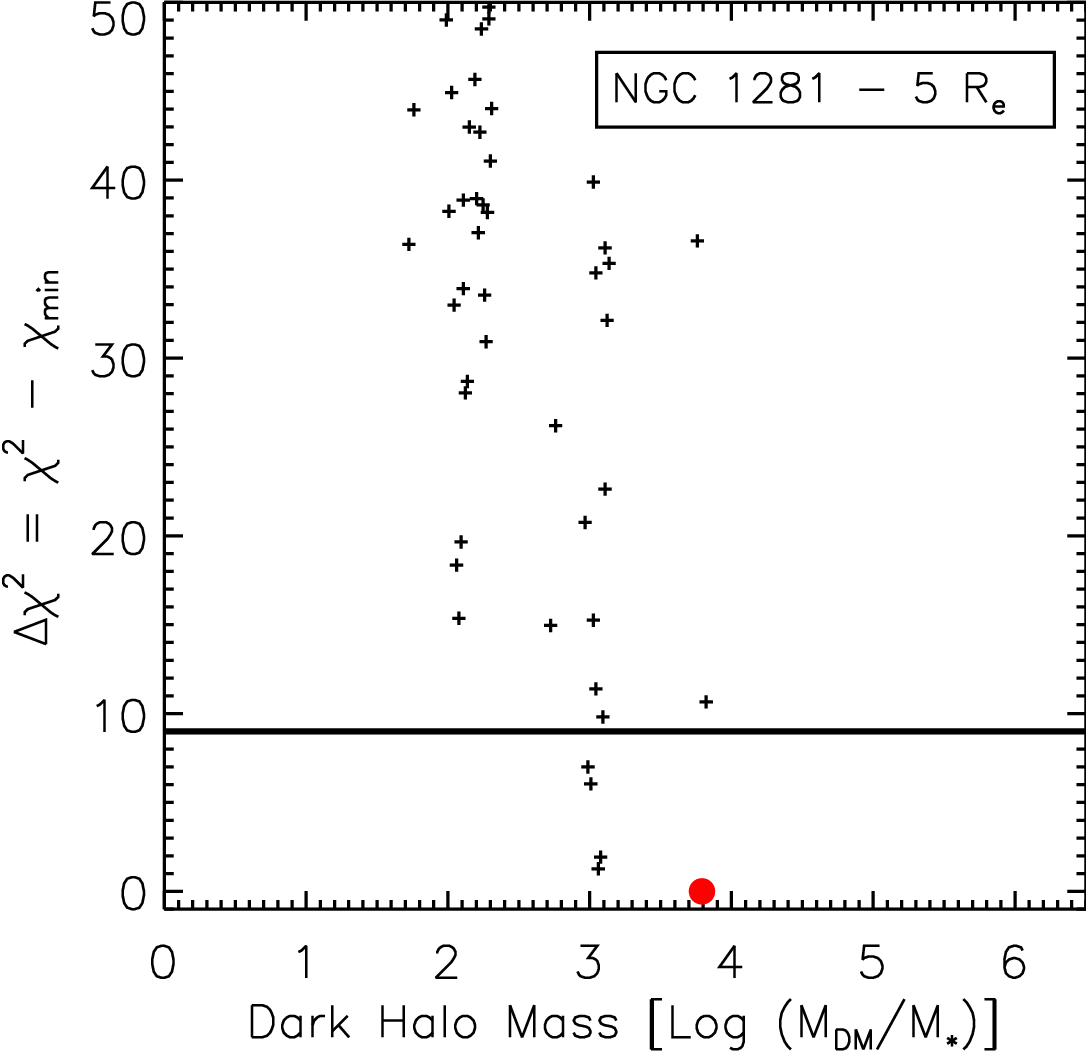}
		\vspace{4ex}
		\includegraphics[width=.49\linewidth]{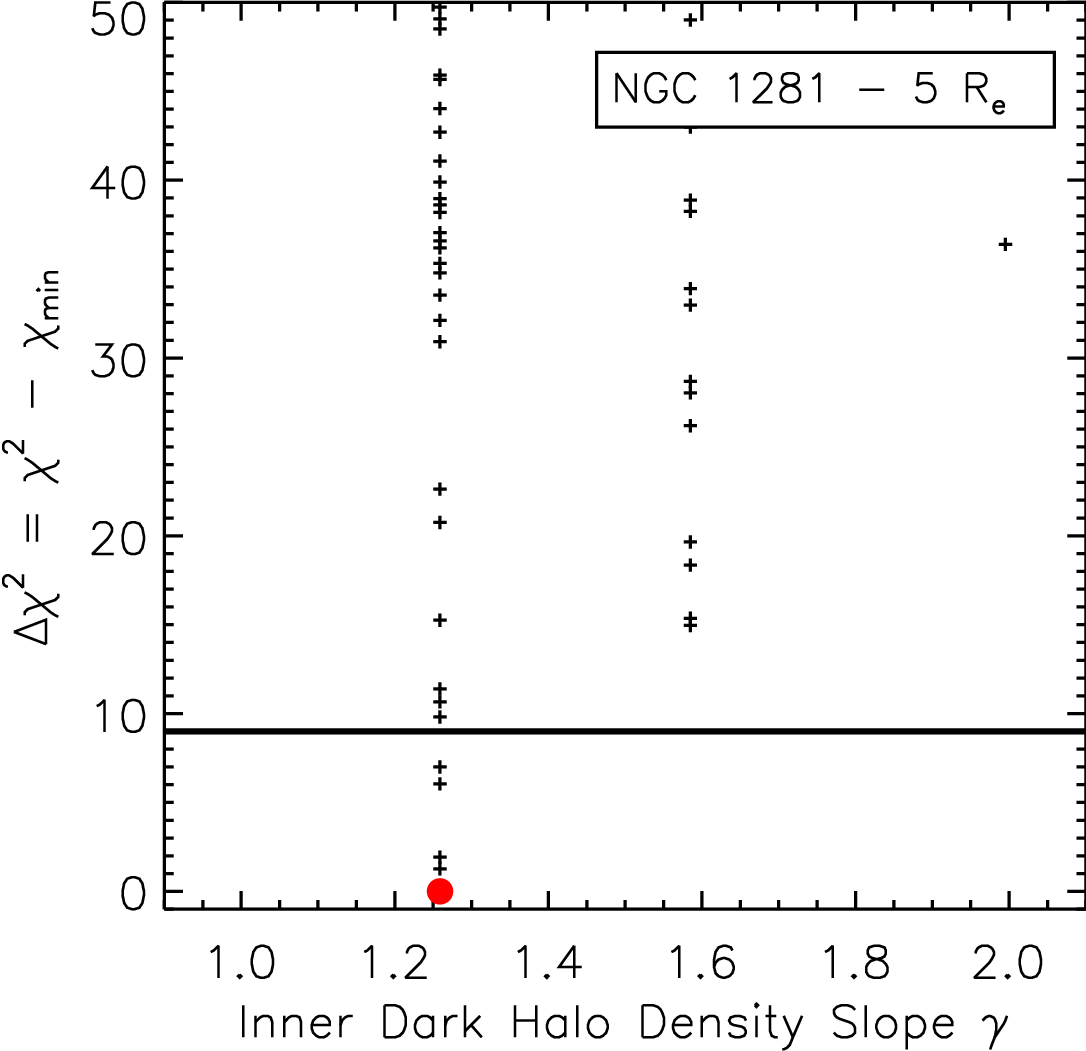}
		\includegraphics[width=.49\linewidth]{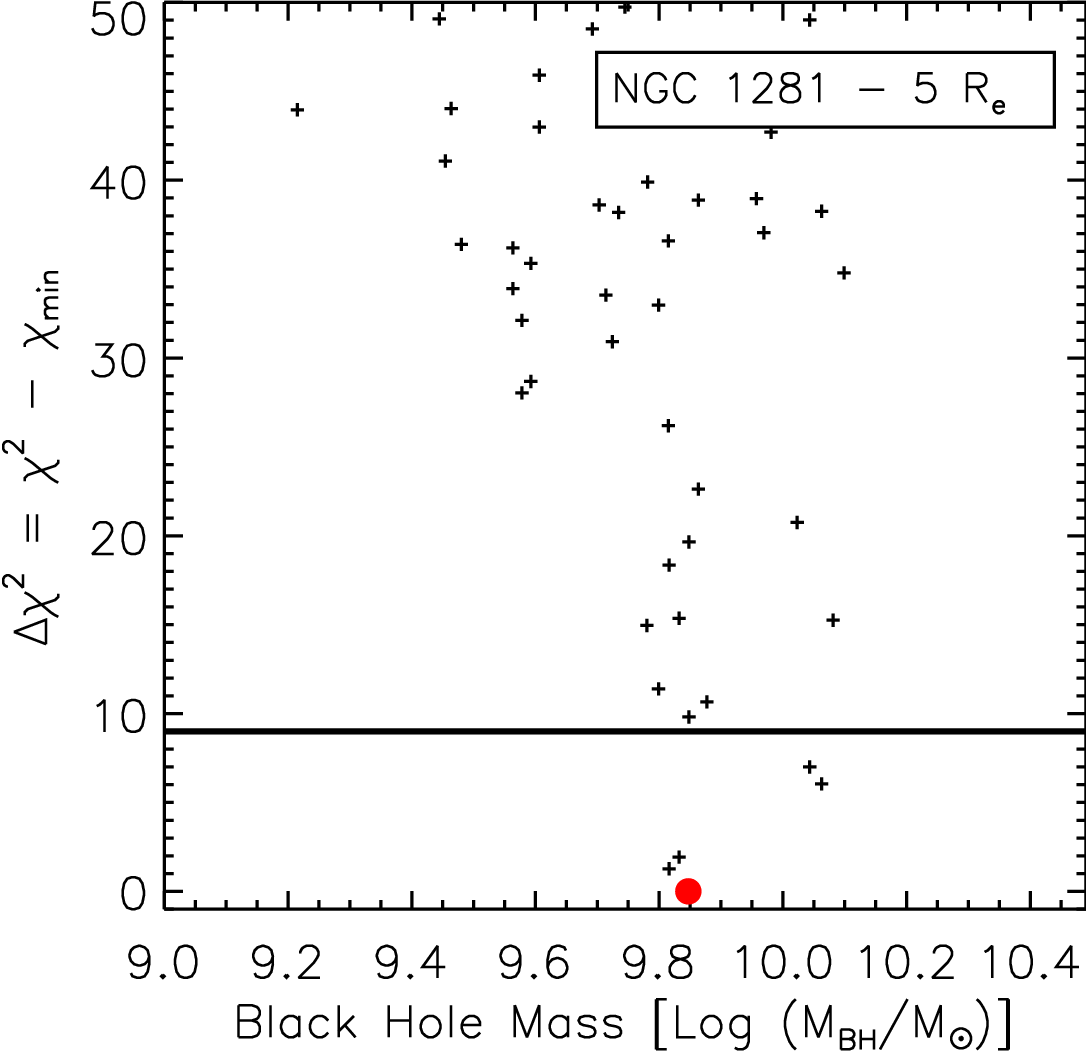}
		\includegraphics[width=.49\linewidth]{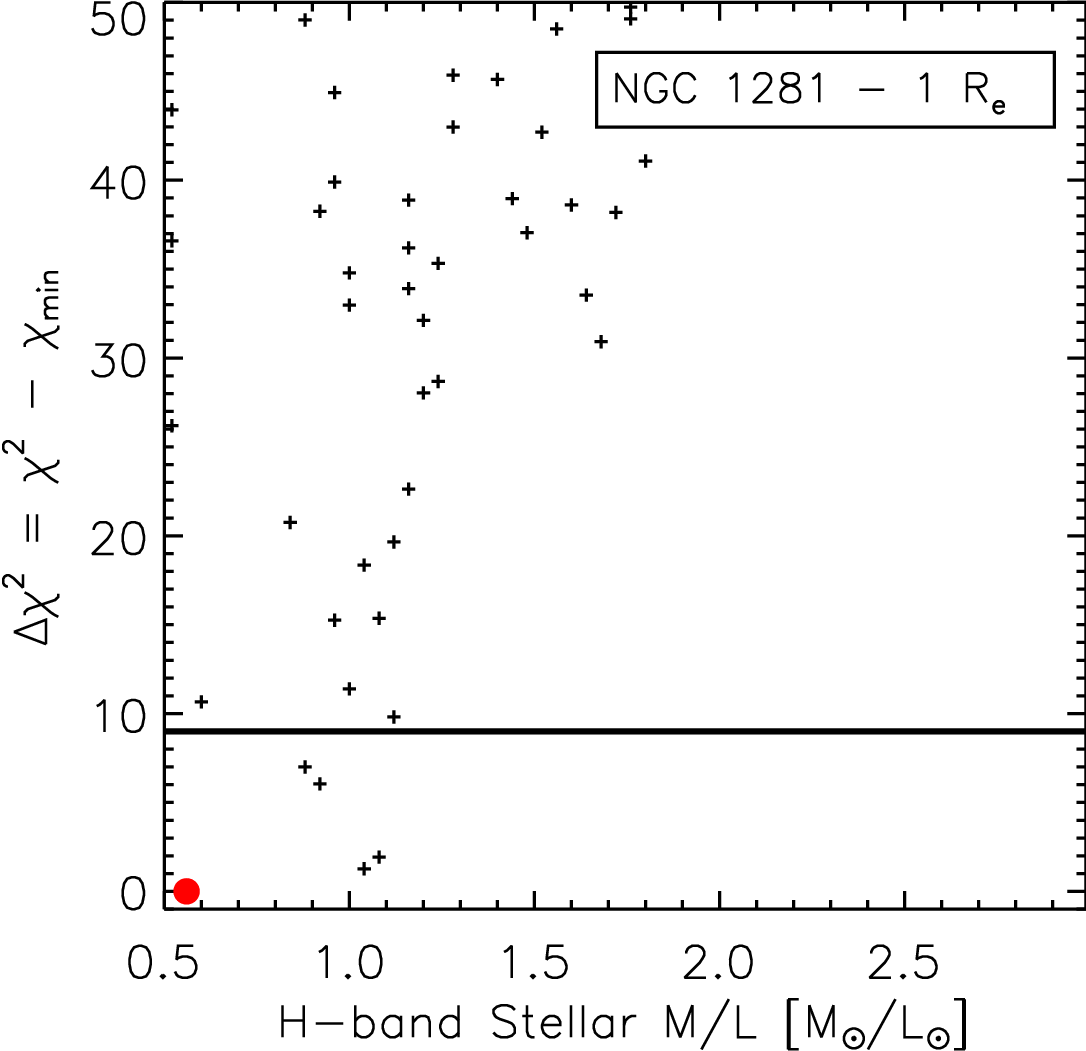}
\caption{Parameter constraints from our orbit-based dynamical models of NGC\,1281's wide-field IFU data, assuming a generalised NFW halo with a varying inner density slope of 1 < $\gamma$ < 2. The plots illustrate the dark halo to stellar mass ratio log($M_{DM}$/\Mstar), the inner dark halo density slope $\gamma$, black hole mass log(\Mbh/$M_{\scriptstyle \odot}$) and stellar mass-to-light ratio in $H$-band. The horizontal line denotes a $\Delta \chi^2$ difference of 9, which corresponds to statistical 3$\sigma$ uncertainties for one degree of freedom, and the red dot corresponds to the overall best-fitting model parameter. The figures endorse our findings of a massive dark halo that, even if parameterised by a generalised NFW profile, leads to questionable results for the black hole mass and stellar M/L in NGC\,1281.}
\label{fig:ngc1281_gnfw}
\end{figure}

Dissipational cosmological simulations advocate that the condensation of baryons in the centre of galaxies alters the central dark matter density distribution. The dark matter halo reacts to this accumulation of baryonic mass with adiabatic contraction \citep{1986ApJ...301...27B}, thereby increasing the central dark matter density profile. To mimic the effect of adiabatic contraction we make use of a generalised NFW profile \citep{1996MNRAS.278..488Z} of the form of:
\begin{equation}
\rho(r) = \frac{\delta_c\;\rho_{c}}{(r/r_s)^\gamma \; (1+ r/r_s)^{3-\gamma}}
\end{equation}
where $\delta_c$ is the characteristic density, $\rho_c = 3H^2/8\pi G$ the critical density, $r_s$ the halo scale radius and $\gamma$ the dark matter density slope. Given our inability to constrain $c_{DM}$, we fix the halo concentration to log($c_{DM}$) = 1 and now examine models with dark matter density slopes of 1 < $\gamma$ < 2. Fig. \hyperref[fig:ngc1281_gnfw]{\ref{fig:ngc1281_gnfw}} summarises our findings, where we display the dark halo to stellar mass ratio log($M_{DM}$/\Mstar), the inner dark halo density slope $\gamma$, black hole mass log(\Mbh/$M_{\scriptstyle \odot}$) and stellar mass-to-light ratio in $H$-band as a function of $\chi^2$, when fitting the full LOSVD. To qualitatively assess if models with an increased density slope would provide more sensible parameter constraints, we adopted a coarse sampling strategy in $\gamma$, black hole mass \Mbh\ and dark halo mass $M_{DM}$. As shown below, an increase in the central density slope draws a similar picture as our fiducial NFW models with a slope of $\gamma$ =1. Our small set of models rules out slopes of $\gamma$ $\ge$ 1.25, and hereby disfavours higher density slopes that could have formed during contraction. However, the results have to be treated with caution, as these models still fail to give comforting estimates for all parameters in the fit, with a stellar M/L that continues to be confined to a range of  0.5 $\le$ \MLstar/\MLsun $\le$ 1, a halo virial mass that exceeds log($M_{DM}$/\Mstar) = 3 and a black hole mass of the order of log(\Mbh/$M_{\scriptstyle \odot}$) = 9.8.\\

Throughout this paper, we have assumed a spherical halo density distribution, but the data also allow for models with a flattened halo. In fact, high-resolution numerical simulations predict triaxial halo density profiles \citep{2002ApJ...574..538J}, if only with a mild flattening of q > 0.7 \citep{2007MNRAS.377...50H}. \cite{2006PASA...23..125K} provide an estimate of the error when averaging an elliptical dark halo mass distribution by a spherical one, showing that the difference well within the scale radius is only at the percentage level for mildly flattened halos. It remains therefore questionable how such a slightly flattened halo, which would still be parameterised by an NFW profile, would be able to resolve the difficulty in recovering the high total enclosed mass while simultaneously providing reasonable estimates for the remaining parameters in the fit.

%---------------------------------------------------------------------
\subsubsection{Data and Uncertainties}
\label{sec:reliability}
%---------------------------------------------------------------------

The detection of a massive dark halo in NGC\,1281 hinges on the reliability of the extracted LOSVD. We have therefore tested the robustness of the LOSVD extensively, by allowing for variations in the width of the spectral masks, the inclusion of multiplicative Legendre polynomials, the use of the MILES stellar library, as well as by constraining the Indo-U.S. stellar library to only a subset of stellar templates that best fits the spectra of the highest S/N bins in the centre. The resulting LOSVDs are consistent within the 2$\sigma$ measurement errors and the mean deviations between the individual measurements amount to 5 \kms\ and 10 \kms\ for $v$ and $\sigma$, respectively, and to 0.02 for $h_3$ and $h_4$. Fitting these individual kinematic data sets then again yields insignificant changes for the recovery of the fitting parameters that are reported in Sec. \hyperref[sec:ngc1281_schwarzschild]{\ref{sec:ngc1281_schwarzschild}}. The parameter constraints for the dark halo, stellar M/L and black hole mass remain largely identical within the statistical 3$\sigma$ confidence intervals, with only small variations in the locations of the best-fitting parameters.
Considering the differences between the various data sets, we have also carried out models where we included systematic errors for our fiducial kinematics (Fig. \hyperref[fig:ngc1281_kin]{\ref{fig:ngc1281_kin}}) of 5 \kms\ and 10 \kms\ for $v$ and $\sigma$. The modelling results show an increase in the 3$\sigma$ upper bound of the stellar M/L from \MLstar\ = 1.1 \MLsun\ to \MLstar = 1.3 \MLsun, while decreasing the 3$\sigma$ lower bound of the dark halo from log($M_{DM}$/$\Mstar$) = 3.1 to log($M_{DM}$/$\Mstar$) = 2.7. Still, the black hole remains over-massive (log(\Mbh/$M_{\scriptstyle \odot} \ge$ 9.8)) and the dark halo and stellar M/L are only at the edge of being consistent with expectations from the stellar-to-halo mass relation and our analysis of the stellar populations in Sec \hyperref[sec:ngc1281_imf]{\ref{sec:ngc1281_imf}}. Since $v$ and $\sigma$ are the dominant driver of the $\chi^2$, we have omitted to include systematic errors for $h_3$ and $h_4$ which are unlikely to significantly decrease the dark halo to stellar mass ratio further.

We would also like to emphasise that, while the remote measurements are the driver of the massive dark halo detection, the dispersion measurements of several of the outermost bins drop below the instrumental resolution. 
To judge the influence of these measurements on our parameter estimation, we have carried out additional tests where we excluded these bins from the fitting process. Given the small number of bins that are affected, though, we could not observe a change in the parameter constraints of these test models. We obtained identical values for the stellar and dark mass in NGC\,1281 with respect to our fiducial models, and effectively rule out any bias of our modelling results in terms of the low dispersion measurements slightly below the instrumental resolution.\\

%---------------------------------------------------------------------
\subsubsection{Outlook}
\label{sec:thoughts}
%---------------------------------------------------------------------

Optimally, one would like to have kinematic tracers at much larger distances, to be able to constrain the total dark halo mass and profile of NGC\,1281. Here, we relied on stellar kinematic data with the \ppak\ IFU and a total integration time of 6 hours on source. Clearly, one can push for much longer exposures and hence deeper kinematics, but the velocity dispersion in NGC\,1281 drops rapidly towards the remote regions and we have already started to reach into regimes where the dispersion falls below the \ppak\ instrumental resolution. As an alternative, the \textit{VIRUS-P} spectrograph can be employed, which is mounted on the 2.7 m Harlan J. Smith telescope at McDonald Observatory \citep{2008SPIE.7014E..70H}. \textit{VIRUS-P} consists of larger 4.3\arcsec\ fibres and covers an even larger FOV of 1.7 $\square$\arcsec. More importantly, though, it has a spectral resolution of 5 \AA\ FWHM (i.e. a velocity resolution of $\sigma$ = 120 \kms) across its 4600 - 6800 \AA\ wide spectral window, and would be suitable to obtain reliable stellar kinematic data beyond 5 \Reff. Ultimately, globular cluster kinematics will be helpful to probe the halo at distances where, according to our fiducial models, we expect a dark halo mass of log($M_{DM}$/$M_{\scriptstyle \odot}$) $\ge$ 13, and thus would provide another test for the validity of a (non-)NFW halo in this object.\\

In this paper, we have adopted the most common dark halo parameterisations, consisting of a NFW profile, a cored-logarithmic profile and of models where mass-follows-light. Different halo parameterisations such as an Einasto profile can, in principle, be adopted, too, and have been shown to provide an even better fit to the profile of simulated dark halos within the $\Lambda$CDM framework \citep[e.g.][]{2006AJ....132.2685M,2009MNRAS.398L..21S,2010MNRAS.402...21N,2014MNRAS.441.3359D}. Given the small differences between both a NFW and Einasto profile, however, such a halo profile appears to be an unlikely recourse to solve the discrepancy between our parameter estimates and those that are promoted e.g. by its stellar populations and the black hole scaling relations. A more promising avenue would be the non-parametric recovery of the total mass profile. In \cite{2006ApJ...641..852V}, for instance, orbit-based dynamical models of the globular cluster M15 have been constructed by including spatial variations in the M/L. A similar scenario with a single M/L slope provided an equally good fit to the LOSVD in our case, but was insufficient to lower the predictions for the dark halo (Sec. \hyperref[sec:ngc1281_ml]{\ref{sec:ngc1281_ml}}). Alternatively, \cite{2013ApJ...763...91J} have also explored non-parametric dark matter distributions. In this way, the total mass profile can be inferred without a priori assumptions of a dark halo profile. Once the total mass profile has been reliably constrained, one can adopt more moderate assumptions for the stellar M/L and/or black hole mass and thus recover the correct dark halo profile of NGC\,1281. We note, though, that the non-parametric recovery of the total mass profile is computationally expensive and compromises need to be made in terms of the spatial information at which the LOSVD and hence the total mass profile will be probed.

%============================= section 5 =============================
\section{Summary}
\label{sec:summary}
%=====================================================================

We have performed a dynamical analysis of the compact, early-type galaxy NGC\,1281. Based on high-resolution imaging and wide-field IFU stellar kinematics, we carried out orbit-based dynamical models to constrain the contribution of stars, black hole and dark matter to its total mass budget. In particular the large-scale kinematics out to 5 \Reff\ ($\simeq$ 7\,kpc) provide strong constraints on the dark and luminous mass in this object.\\

According to our models, NGC\,1281 is a dark matter dominated galaxy with a dark matter fraction of 90 per cent to the total enclosed mass within the kinematic extent, irrespective of the adoption of a NFW or CL halo. Assuming that the dark halo profile can be parameterised by a spherical NFW halo, this yields a dark halo mass of 11.5 $\le$ log($M_{DM}$/$M_{\scriptstyle \odot}$) $\le$ 11.8 within the kinematic extent. Moreover, the extrapolated dark halo profile would imply a dark halo virial mass of log($M_{DM}$/$M_{\scriptstyle \odot}$) = 14.1$^{+1.5}_{-0.5}$. This halo mass is typically ascribed to the halo of galaxy clusters and in contrast to findings based on $\Lambda$CDM cosmology and the halo abundance matching mechanism. The massive dark halo, in turn, tightly constrains the stellar mass-to-light ratio in this object. The stellar mass-to-light ratio is predicted to be 0.6 $\le$ \MLstar/\MLsun\ $\le$ 1.1 in \textit{H}-band, and hence considerably lower than what is expected from our spectral analysis of its stellar populations. This strongly indicates that, even if a NFW halo provides a good fit to the LOSVD and successfully recovers the total enclosed mass within the kinematics bounds, NFW might not be a good approximation of the dark halo profile of this galaxy. 

A spatial variation in the stellar mass-to-light ratio has been considered. Models with a positive slope provide a good fit to the kinematics, but they cannot account for a significant fraction of total enclosed mass due to the rapid decline of the stellar surface brightness. In addition, the positive slopes predict a variation in the stellar M/L which should translate into a strong variation in colours. The lack of such a colour gradient in NGC\,1281, however, endorses the employment of a constant \MLstar.

Adopting a stellar M/L of 1.7 \MLsun\ decreases the dark halo contribution, but provides a significantly worse fit to the stellar kinematics and is neither able reconcile the dark halo mass with those predicted within the $\Lambda$CDM framework nor the stellar M/L with the values promoted by its stellar populations.

Finally, the models predict a SMBH mass of log(\Mbh/$M_{\scriptstyle \odot}$) = $10.0^{+0.1}_{-0.2}$, but the black hole sphere of influence is not resolved by our observations. The detection is merely a result of the unrealistically low stellar mass-to-light ratio, which propagates inwards and predicts the presence of a massive SMBH to be able to explain the central dynamical mass. Fixing the black hole mass according to predictions of the black hole scaling relations naturally increases the stellar M/L and hence decreases the dark halo mass. However, these models also fail to recover the LOSVD, which corroborates the necessity of a non-NFW halo in this object.

%=====================================================================
\section*{Acknowledgements}
A.\,Y. acknowledges support from the International Max Planck Research School for Astronomy \& Cosmic Physics at the University of Heidelberg and the Heidelberg Graduate School of Fundamental Physics. J.\,L.\,W. has been supported by an NSF Astronomy and Astrophysics Postdoctoral Fellowship under Award No. 1102845. K.\,G{\"u}. acknowledges support from the Theodore Dunham, Jr. Grant of the Fund for Astrophysical Research. The data presented here is based on observations collected at the Centro Astronómico Hispano Alemán (CAHA) at Calar Alto, operated jointly by the Max Planck Institute for Astronomy and the Instituto de Astrofísica de Andalucía (CSIC). This research also based on observations made with the NASA/ESA \textit{Hubble Space Telescope}. The observations are associated with program \#13050. Support for program \#13050 was provided by NASA through a grant from the Space Telescope Science Institute, which is operated by the Association of Universities for Research in Astronomy, Inc., under NASA contract NAS 5-26555. This research has made use of the NASA/IPAC Extragalactic Database (NED) which is operated by the Jet Propulsion Laboratory, California Institute of Technology, under contract with the National Aeronautics and Space Administration, and the SDSS DR8. Funding for the SDSS and SDSS-II has been provided by the Alfred P. Sloan Foundation, the Participating Institutions, the National Science Foundation, the U.S. Department of Energy, the National Aeronautics and Space Administration, the Japanese Monbukagakusho, the Max Planck Society, and the Higher Education Funding Council for England. The SDSS Web Site is http://www.sdss.org/.
The SDSS is managed by the Astrophysical Research Consortium for the Participating Institutions. The Participating Institutions are the American Museum of Natural History, Astrophysical Institute Potsdam, University of Basel, University of Cambridge, Case Western Reserve University, University of Chicago, Drexel University, Fermilab, the Institute for Advanced Study, the Japan Participation Group, Johns Hopkins University, the Joint Institute for Nuclear Astrophysics, the Kavli Institute for Particle Astrophysics and Cosmology, the Korean Scientist Group, the Chinese Academy of Sciences (LAMOST), Los Alamos National Laboratory, the Max-Planck-Institute for Astronomy (MPIA), the Max-Planck-Institute for Astrophysics (MPA), New Mexico State University, Ohio State University, University of Pittsburgh, University of Portsmouth, Princeton University, the United States Naval Observatory, and the University of Washington.%=====================================================================

%=====================================================================
% FIGURES
%=====================================================================

%=====================================================================
% APPENDICES
%=====================================================================
%
%=====================================================================

%=====================================================================

%=====================================================================
% REFERENCES
%=====================================================================
%\bibliographystyle{mn2e_fix}
\bibliographystyle{yahapj}
\bibliography{mn2e}

%\bsp

\label{lastpage}

\end{document}